\documentclass[aps,prl,reprint,superscriptaddress]{revtex4-2}
\usepackage{graphicx}
\usepackage{amsmath}
\usepackage{xcolor}

\begin{document}


\title{Dense granular rheology from fluctuations}

\author{Benjamin M. Alessio}
\email{balessio@stanford.edu}
\affiliation{Department of Mechanical Engineering, Stanford University, Stanford, CA 94305, USA}

\author{Matthew R. Edwards}
\email{mredwards@stanford.edu}
\affiliation{Department of Mechanical Engineering, Stanford University, Stanford, CA 94305, USA}

\author{Ching-Yao Lai}
\email{cyaolai@stanford.edu}
\affiliation{Department of Geophysics, Stanford University, Stanford, CA 94305, USA}

\date{\today}

\begin{abstract}

A governing equation for the granular temperature of flows past the jamming point is proposed and validated term-by-term with discrete element simulation. We measure and construct models for transport coefficients that do not diverge near the jamming fraction. As a limiting case of our equation, we derive the granular fluidity equation and find disagreement between the preexisting parametrizations of nonlocal granular fluidity, extended kinetic theory, and our discrete element simulations that is resolved by our continuum model.

\end{abstract}

\maketitle 

Dense granular flows have resisted decades of modeling efforts. A complete description linking microphysical, grain-to-grain interactions with macrophysical transport coefficients remains an outstanding classical physics problem~\cite{kamrin2024advances}. Many of the challenges can be traced to the fact that grains dissipate energy during collisions~\cite{wolf1998dissipation}. As a consequence, dilute flows amenable to highly descriptive, physically-grounded kinetic theories~\cite{garzo1999dense,brilliantov2010kinetic}, if somehow not complicated by ever-present interstitial fluids~\cite{pahtz2019local}, tend to condense into a phase of interest for practical applications in geophysics~\cite{jerolmack2019viewing} and industry~\cite{govender2016granular}. This dense phase is typically considered abundant in fluctuating motions that arise from geometric incompatibility of affine motions--a grain moving precisely with the mean flow will inevitably be diverted by its neighbors. Non-affine motions are the root of viscous resistance to flow for any fluid~\cite{zaccone2023general}, and therefore a granular rheology should, explicitly or implicitly, characterize fluctuations. The absence of Brownian motion means that granular fluctuations are generated entirely by macroscopic shearing, and therefore the task of deducing transport coefficients from fluctuations comes with difficulties not unlike the same task for turbulent flows~\cite{savage1992instability,radjai2002turbulentlike,miller2013eddy}. 

Development of predictive models for dense granular flows has largely been a matter of phenomenological extensions to idealized models, guided by experiments and discrete element method (DEM) simulations~\cite{cundall1979discrete,luding2008introduction}. One such idealized model is the empirical inertial rheology~\cite{jop2006constitutive}, which relates the shear-normal stress ratio $\mu$ to the shear rate $\dot\gamma$, pressure $p$, mean grain diameter $d$, and individual grain density $\rho$ using the dimensionless inertial number $I=\dot\gamma d\sqrt{\rho/p}$~\cite{da2005rheophysics} and features a critical value $\mu=\mu_s$ below which the material does not yield. To account for the widely observed yielding behavior $\mu<\mu_s$ when the flow is heterogeneous in strain rate (particularly over a length scale an order of magnitude greater than the grain size), several extensions have been proposed~\cite{kamrin2024advances}, each with the feature of a reacting, transporting state variable that captures nonlocal effects. The state variable is typically interpreted as a phase descriptor in a continuous change between jammed and flowing states, and can be referred to as the ``fluidity''~\cite{aranson2002continuum,bocquet2009kinetic,bouzid2015non}. In many of these models, the transport of this state variable takes the form of simple diffusion in an ordinary differential equation~\cite{aranson2002continuum,bazant2006spot,kamrin2012nonlocal,bouzid2013nonlocal,henann2013predictive}, although integral equation models have also been proposed~\cite{pouliquen2009non,rognon2015long,dsouza2020non}. It is clear that the grain-scale fluctuations play an important role in fluidizing the material~\cite{dijksman2011jamming,wortel2016criticality,zhang2017microscopic,degiuli2017friction,gautam2025micromechanical}. Recently, it was hypothesized that the rescaled granular temperature $\Theta = \rho T/p$, a measure of fluctuations, can be used to cast the uncollapsed $\mu(I)$ relation including nonlocal effects into a one-to-one relation $\mu\Theta^{a_\mu}=f(I)$~\cite{kim2020power}. Absent a governing equation for $T$ (or $\Theta$), this model is not predictive. Irmer et al.~\cite{irmer2025granular} found exponential decay-like spatial profiles of $\Theta$ in wall-vibrated shear flows and suggested that a heat equation could be appropriate, although as we will show, their model does not generalize to different flow scenarios.

Another such idealized model is the extended kinetic theory (EKT) with a phenomenological treatment of the post-collision velocity correlations~\cite{losert2000particle,bocquet2001granular,jenkins2006dense,jenkins2007dense,berzi2015different,berzi2018fluidity,vescovi2024extended,berzi2024granular}. In a recent breakthrough, Berzi~\cite{berzi2024granular} identified that, for the limiting case of flows below the critical packing fraction, where rate-independent contact stresses are absent, EKT provides explicit expressions for the fitting parameters of the inertial rheology and the nonlocal granular fluidity (NGF~\cite{kamrin2024advances}) model. However, fluidity models were originally motivated by yield stress materials~\cite{goyon2008spatial,bocquet2009kinetic}, corresponding to granular materials above the critical packing fraction where rate-independent stresses cannot be ignored and EKT becomes inaccurate~\cite{debnath2022different}. The fluidity models have been quantitatively validated in situations where rate-independent stresses clearly dominate~\cite{alessio2025supplemental,zhang2017microscopic,fazelpour2022effect,irmer2025granular}. Several further extensions of the kinetic theory have been proposed that account for rate-independent stresses~\cite{johnson1987frictional,savage1998analyses,campbell2002granular,berzi2011constitutive,berzi2011surface,chialvo2012bridging,chialvo2013modified,berzi2015steady,vescovi2016merging,duan2019new,berzi2019erodible,berzi2020extended,chassagne2023frictional} and predict features such as the energy spectrum~\cite{saitoh2016anomalous,saitoh2017anisotropic}. Still, these further extensions have neither been connected to the fluidity models nor been directly tested, for dense flows of frictional grains above the critical packing fraction, against DEM or experiments.

\begin{figure}
\includegraphics[width=0.49\textwidth]{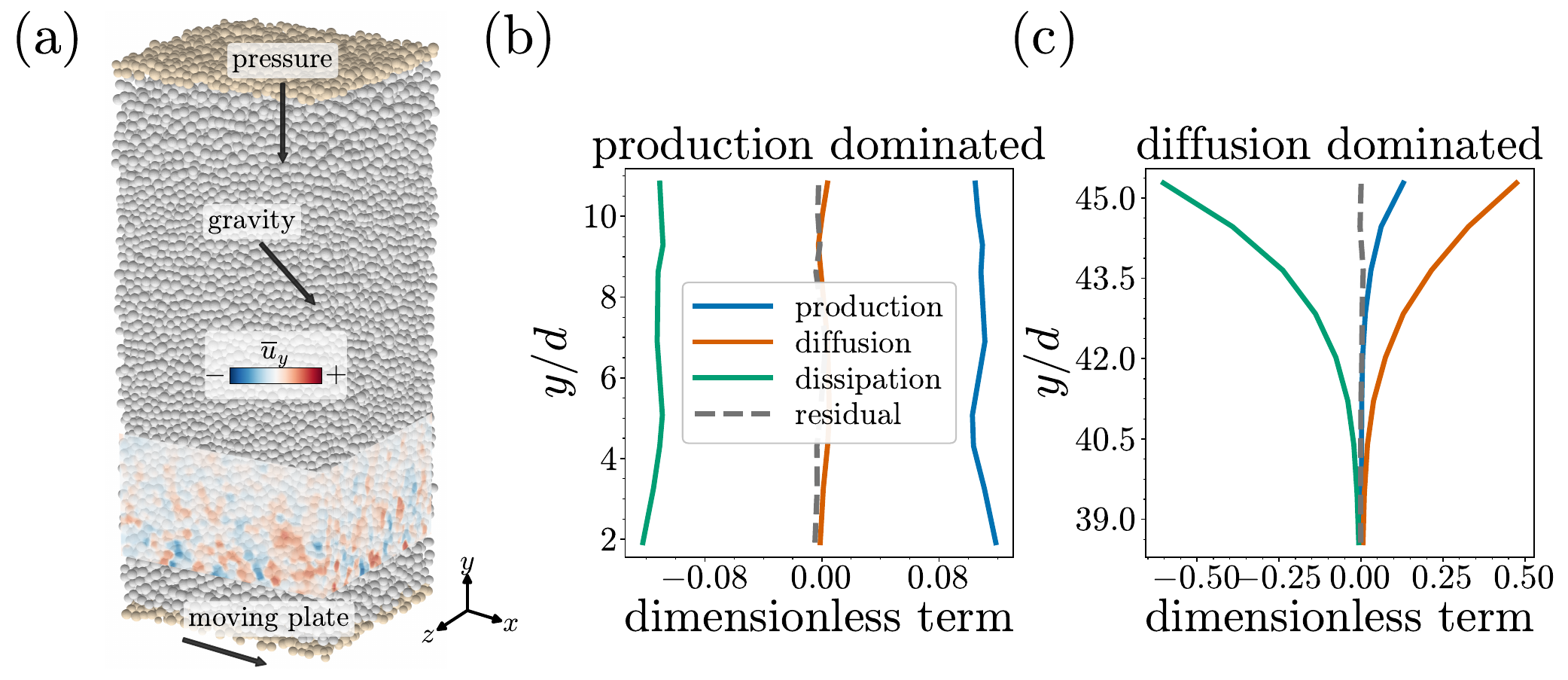}
\caption{\label{fig:schematic}Illustration of a DEM simulation, coarse-graining, and ensemble-averaging. (a) A shear cell is driven by plate motion and the horizontal component of gravity, and has pressure controlled by the top plate and the vertical component of gravity. A semi-translucent planar band shows an example instantaneous $\overline{u}_y$ field, coarse-grained with length scale $d/2$. For grain elastic wave speed $u_w$, $O(\overline{u}_y)\ll u_w$. (b,c) Ensemble averaged spatial profiles, assuming homogeneity in $x$ and $z$, of each term in Eq.~\ref{eqn:energy} are shown for simple shear (b) and shear with gravity (c). Each term is nondimensionalized by the $y$-average of $\chi$.}
\end{figure}

In this Letter we close the system of equations for a granular temperature-based steady rheology of dense granular flows above the critical volume fraction and perform extensive DEM simulations for direct validation. We derive the granular fluidity equation from linearization of our equation, and find the distinct parameterizations of the fluidity equation, from the original NGF and from EKT, as well as a recently proposed heat equation for the granular temperature, possess qualitative differences with our simulation data that are resolved by our model. 

Various forms of the granular temperature evolution equation have been derived~\cite{pahtz2015fluctuation,berzi2020extended}, generally from a decomposition where a field $f=\langle f\rangle + f'$ has an ensemble mean $\langle f\rangle$ and a fluctuation $f'$ such that $\langle f'\rangle=0$~\cite{j2007statistical}. We notate grain-scale coarse-grained quantities, which are continuous and fluctuating~\cite{goldhirsch2010stress} (see also Supplemental Information (SI)~\cite{alessio2025supplemental}), separately from the ensemble mean. Defining grain-scale, coarse-grained fields for density $\varrho$, velocity $u_i$, and contact stress tensor $\sigma_{ij}^\text{con}$, and ensemble-averaged fields for stress tensor $\sigma_{ij}=\langle\sigma_{ij}^\text{con}-\varrho u_i'u_j'\rangle$, fluctuating energy flux $Q_j=\langle\sigma_{ij}^\text{con}u_i'-\frac12\varrho u_i'u_j'u_i'\rangle$, dissipation rate $\chi=\langle\sigma_{ij}^\text{con}\partial u_i/\partial x_j\rangle$, and granular temperature $T=\langle u_i'u_i'\rangle/2$, the steady state of the granular temperature equation satisfies~\cite{pahtz2015fluctuation}
\begin{equation}
\label{eqn:energy}
\begin{split}
\sigma_{ij} \frac{\partial \langle u_i\rangle}{\partial x_j} - \frac{\partial Q_j}{\partial x_j} - \chi = 0.
\end{split}
\end{equation}
The first, second, and third terms on the left-hand side are shear production, diffusion, and dissipation. The unclosed quantities are $\sigma_{ij}$, $Q_j$, and $\chi$, and we note that closure (modeling in terms of known quantities) of $\sigma_{ij}$ is required to solve the momentum equation for $\langle u_i\rangle$~\cite{alessio2025supplemental}. In general, $\sigma_{ij}$ and $Q_j$ have contributions from grain-grain interaction forces and velocity fluctuations (see~\cite{berzi2024granular} for discussion regarding kinetic, collisional, and rate-independent stresses). Our separation of production and dissipation is somewhat arbitrary. The production of $T$ alone, not considering other forms of energy, can only come from $-\langle\varrho u_i'u_j'\rangle\partial\langle u_i\rangle/\partial x_j$~\cite{johnson1987frictional,berzi2011constitutive}. Defining the production in Eq.~\ref{eqn:energy} that way would require redefining $\chi$ as $\langle\sigma_{ij}^{\text{con}\prime}\partial u_i'/\partial x_j\rangle$, another standard form~\cite{pope2001turbulent,chassagne2023frictional}. We instead minimize the number of unclosed terms by considering all contributions of stress together in $\sigma_{ij}$~\cite{savage1998analyses,pahtz2015fluctuation}. The dissipation is then understood~\cite{pahtz2015fluctuation} to represent the conversion of translational kinetic energy into rotational, potential, or irreversible thermal energy (routinely measured in DEM~\cite{itascaPFC,rangel2023experimental} and modeled~\cite{lorenz2019theory}). We note that this kinematic form of dissipation can in general be positive or negative owing to the rotational and potential contributions, but at steady state two-stage irreversibility~\cite{wiebicke2024simple} requires it be positive. 

\begin{figure}
\includegraphics[width=0.49\textwidth]{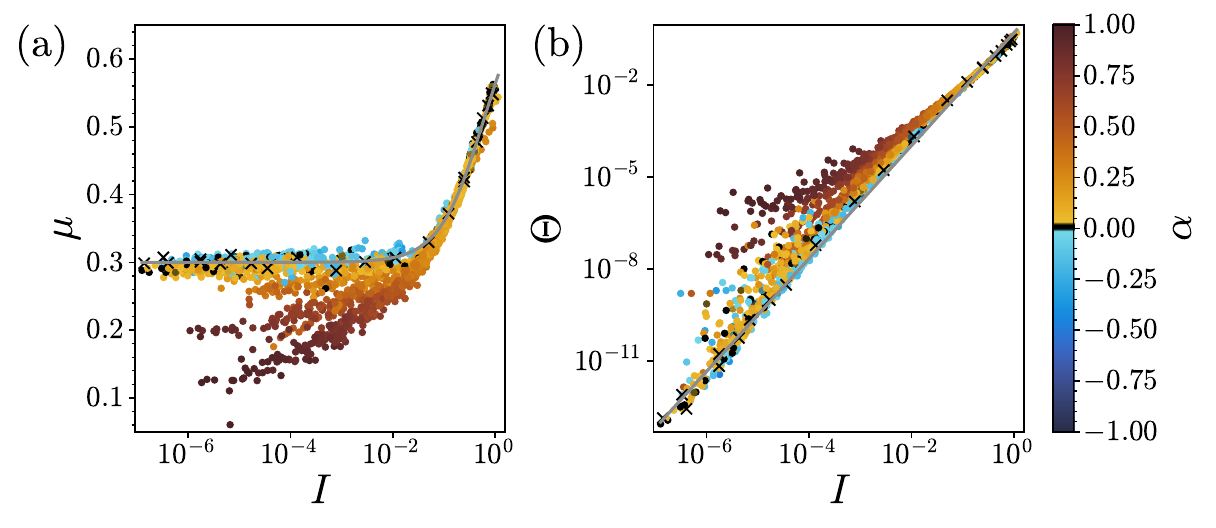}
\caption{\label{fig:mu_and_theta}Friction $\mu$ and dimensionless granular temperature $\Theta$ from DEM vs. inertial number $I$. The deviations from the $\mu(I)$ hypothesis result from the influence of the independent granular temperature field. Data for homogeneous shear tests are shown in crosses with gray fit lines for $\mu_0(I)$ and $\Theta_0(I)$, and all points are colored by relative diffusion $\alpha$ according to a custom diverging map. See SI~\cite{alessio2025supplemental} for fitting procedures and functional forms.}
\end{figure}

\begin{figure*}
\includegraphics[width=0.99\textwidth]{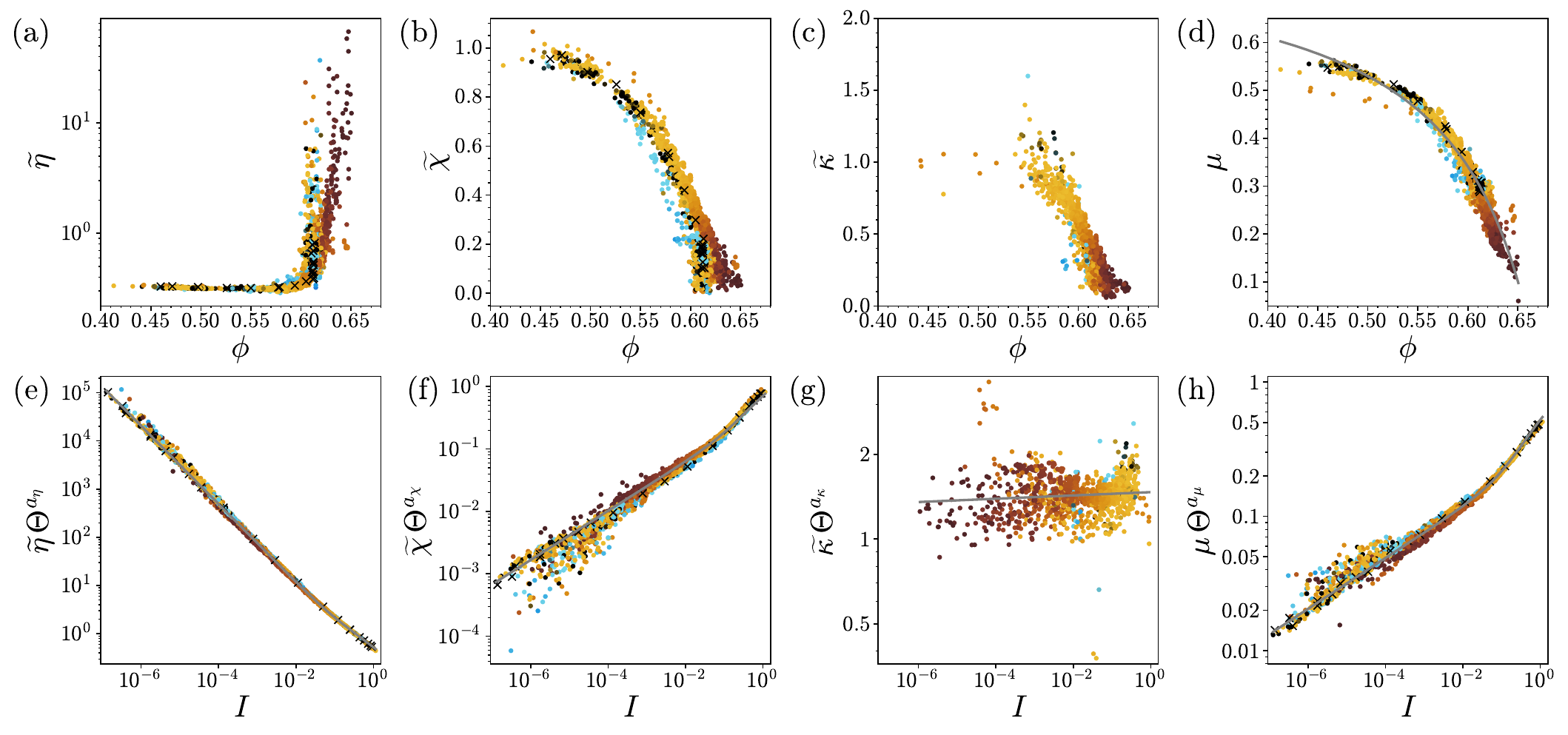}
\caption{\label{fig:scalings}Scalings from DEM data, colored by relative diffusion $\alpha$ (colorbar in Fig.~\ref{fig:mu_and_theta}). Dimensionless viscosity $\tilde\eta$ (a,e), dissipation rate $\tilde\chi$ (b,f), conductivity $\tilde\kappa$ (c,g), and friction $\mu$ (d,h) are plotted against packing fraction $\phi$ (a-d) and, after rescaling by powers of dimensionless temperature $\Theta$, against inertial number $I$ (e-h). Circles correspond to individual $y$-locations of heterogeneous profiles, and crosses to averages over the homogeneous shear profiles. The gray lines in (d,g) are curve fits, and in (e,f,h) are predictions using the fits from Fig.~\ref{fig:mu_and_theta}. See SI~\cite{alessio2025supplemental} for fitting procedures.}
\end{figure*}

Using extensive DEM simulations for dense, unidirectional flows of grains with Hertz-Mindlin contact forces~\cite{barber2000contact} and Tsuji damping~\cite{tsuji1992lagrangian}, we compute each term in Eq.~\ref{eqn:energy} by coarse-graining~\cite{zhang2010coarse,goldhirsch2010stress,weinhart2013coarse} and ensemble-averaging. See SI~\cite{alessio2025supplemental} for simulation details, including our implementation~\cite{hu2019taichi}. Microphysical expressions of the unclosed terms, appropriate for analyses of DEM simulations, have been previously derived~\cite{pahtz2015fluctuation}. A schematic of our DEM model and examples of an instantaneous fluctuating field and ensemble-averaged fields sampled from our simulations are shown in Fig.~\ref{fig:schematic}. Panel (b) shows ensemble-averaged spatial distributions of each term in Eq.~\ref{eqn:energy} for simple shear, where production balances dissipation in a homogeneous way. For gravity driven flow, Fig.~\ref{fig:schematic}c shows that the combined effect of production and diffusion is balanced by dissipation. The small residuals~\cite{alessio2025supplemental} indicate that Eq.~\ref{eqn:energy} is appropriate for our system. In our simulations, we have $\Theta\ll1$, indicating that the rate-independent stresses dominate.

Fig.~\ref{fig:mu_and_theta} plots $\mu$ and $\Theta$ vs. $I$ from our data (using $p\equiv\sigma_{yy}$) to demonstrate that the diffusion term in Eq.~\ref{eqn:energy} is responsible for deviations from the homogeneous $\mu(I)$ rheology. We define relative diffusion $\alpha=\frac{dQ_y}{dy}/\chi$, where a material point with positive $\alpha$ receives $T$ from its surroundings and a point with negative $\alpha$ expels it. We introduce fitted relations using the inertial rheology parameters $\mu_s$, $\mu_2$ and $I^\star$ for $\mu_0(I)=\mu_s+(\mu_2-\mu_s)/(1+I^\star/I)$~\cite{jop2006constitutive} and a simple power law $\Theta_0(I)=A_\Theta I^{n_\Theta}$ to approximate the homogeneous state (see SI~\cite{alessio2025supplemental} for fitting details).

To solve Eq.~\ref{eqn:energy}, it is necessary to apply closure models to $\sigma_{ij}$, $Q_i$, and $\chi$. A central finding from experiments~\cite{poon2023microscopic}, DEM simulations~\cite{zhang2017microscopic}, and nonlocal theories~\cite{dsouza2020non} is that, even for jammed flows, the rheology is rate-dependent with a coefficient of viscosity dependent on $T$ and the packing fraction $\phi=\langle\varrho\rangle/\rho$~\cite{zhang2017microscopic,poon2023microscopic}. This was first theorized for granular flows by Savage~\cite{savage1998analyses} who generalized the original argument of Hibler~\cite{hibler1977viscous} that local directional fluctuations of the strain rate $\dot\epsilon_{ij}$ in a plastically deforming material manifest closures of $\sigma_{ij}$, $Q_i$, and $\chi$ analogous to those of a fluid where (for unidirectional flow in $x$ varying along $y$) $\sigma_{xy}=\eta\dot\gamma$ and $Q_y=-\kappa\partial T/\partial y$. In Savage's theory, the transport coefficients $\eta$, $\kappa$, and $\chi$ are each rescaled by $p$ and $\sqrt T/d$ into dimensionless quantities that are functions of $\phi$, which we define as $\tilde\eta=\eta\sqrt T/pd$, $\tilde\kappa=\kappa\sqrt T/pd$, and $\tilde\chi=\chi d/p\sqrt T$. This nondimensionalization has been used to analyze experiments~\cite{losert2000particle} for all three coefficients, and is supported by DEM simulations~\cite{zhang2017microscopic} and experiments~\cite{poon2023microscopic} for the fluidity $g\equiv p/\eta$. However, the curve $gd/\sqrt T$~vs.~$\phi$ loses accuracy as packing fraction increases beyond the jamming fraction~\cite{kim2020power,gaume2020microscopic}, and may be improved by hypothesizing power law dependence on $\Theta$ and $I$ instead of $\phi$~\cite{kim2020power}. The appreciable spread in the measurement of $gd/\sqrt T$ vs $\phi$ (inverse $y$-axis of Fig.~\ref{fig:scalings}a) has been observed and attributed to additional $\Theta$ dependence~\cite{kim2020power}, crystallization~\cite{fazelpour2022effect}, flow geometry~\cite{robinson2021examination} (although the flow was not Lagrangian-steady), and boundary conditions influencing the nature of the fluctuations~\cite{gaume2020microscopic,poon2023microscopic}. We speculate that the non-universality of the distribution function of fluctuating motions, whose form apparently depends on $\alpha$ (see SI~\cite{alessio2025supplemental}) and deviates from the Gaussian distribution previously assumed~\cite{savage1998analyses}, weakens the accuracy of Savage's scalings. In any case the relation $gd/\sqrt T = f(\phi)$ cannot be considered a \textit{microscopic definition} (as previously claimed~\cite{zhang2017microscopic}) so much as an approximate constitutive model.

\begin{figure}
\includegraphics[width=0.49\textwidth]{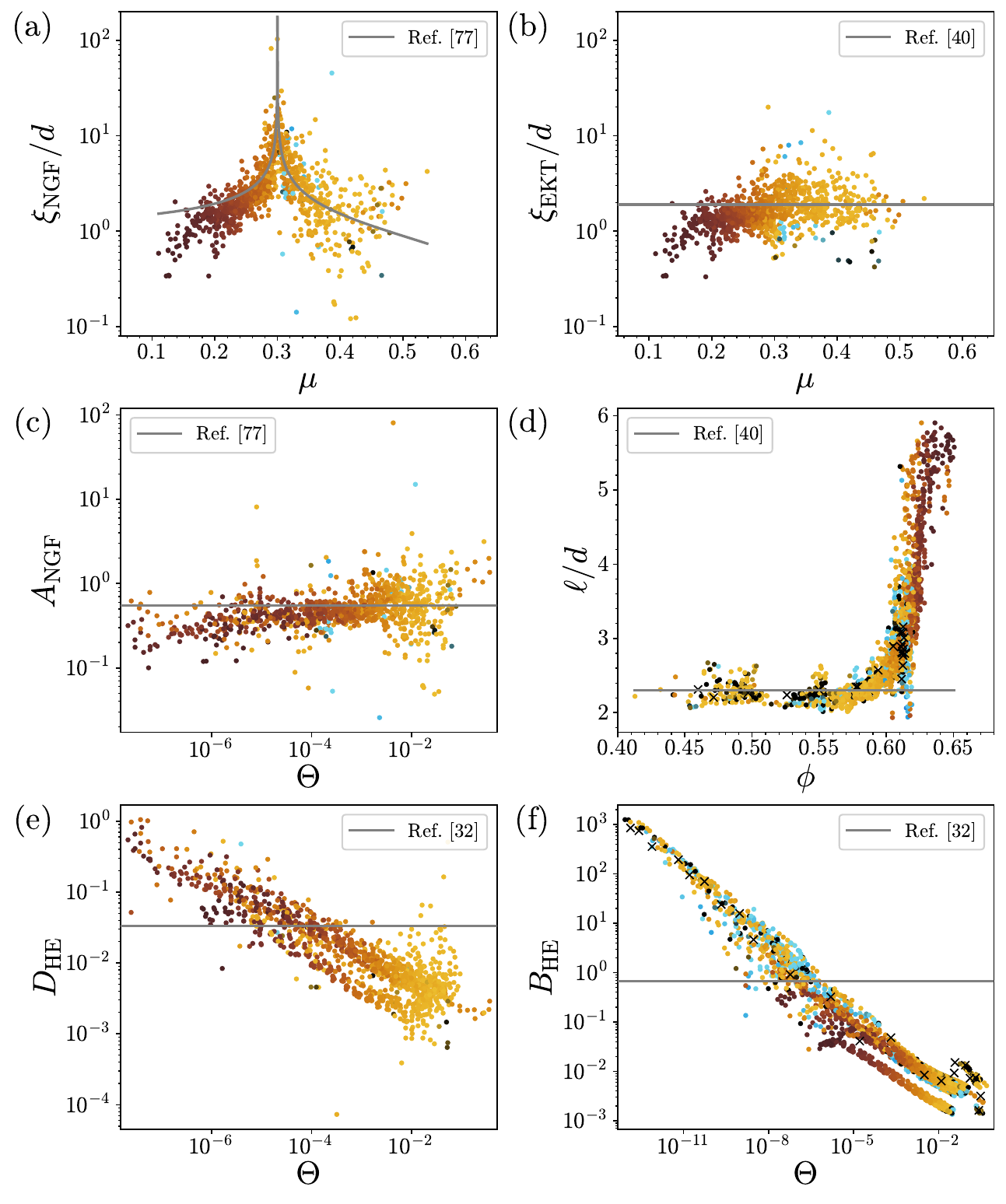}
\caption{\label{fig:params_other_models}DEM measurements of parameters relevant to existing constitutive models for dense granular flows, colored by relative diffusion $\alpha$ (colorbar in Fig.~\ref{fig:mu_and_theta}) with previously hypothesized forms~\cite{kamrin2014nonlocal,berzi2024granular,irmer2025granular} as gray lines. (a) The cooperativity length $\xi_\text{NGF}$ from NGF~\cite{kamrin2014nonlocal} vs. friction $\mu$. (b) The cooperativity length $\xi_\text{EKT}$ from EKT~\cite{berzi2024granular} vs. friction $\mu$. (c) The nonlocal amplitude $A_\text{NGF}$ vs. dimensionless granular temperature $\Theta$. (d) The correlation length $\ell$ scaled by grain size $d$ vs. packing fraction $\phi$. (e,f) The diffusion $D_\text{HE}$ and dissipation $B_\text{HE}$ coefficients from Ref.~\cite{irmer2025granular}. See SI~\cite{alessio2025supplemental} for fitting details.}
\end{figure}

Neither DEM nor experimental measurements of $\tilde\kappa$ and $\tilde\chi$ exist sufficiently extensive to investigate constitutive hypotheses for dense granular flows. To address this gap, we use our extensive DEM simulations to obtain these coefficients. In Fig.~\ref{fig:scalings} we plot the dimensionless coefficients in two manners. First, we show in panels (a-d) that Savage's scalings capture the orders of magnitude of variations in the transport coefficients, which are well-collapsed compared to $\eta$, $\kappa$, and $\chi$ vs. $\phi$ in panels (a-c) of Fig.~\ref{fig:unscaled}. Second, we extend the recently proposed ansatz $\mu\Theta^{a_\mu}=f_\mu(I)$~\cite{kim2020power} to the transport coefficients as $\tilde\eta \Theta^{a_\eta}=f_\eta(I)$, $\tilde\kappa\Theta^{a_\kappa}=f_\kappa(I)$ and $\tilde\chi\Theta^{a_\chi}=f_\chi(I)$ (where we measure the parameters $a_\mu$, $a_\eta$, $a_\kappa$, and $a_\chi$). We plot these scalings in panels (e-h), observing alleviation from the latter (panels a and b) of the singular and diverging behavior near the jamming fraction ($\phi\approx0.61$) with the former. We observe $f_\kappa\approx\text{const.}$ See SI~\cite{alessio2025supplemental} for the details of the exponent-fitting in the power law ansatzes. Some spread in $\tilde\kappa$ at high $\Theta$, low $\phi$ could be attributed to the emergence of the packing fraction gradient contribution to $Q$~\cite{berzi2024granular}. Each data point in Figs.~\ref{fig:mu_and_theta},~\ref{fig:scalings}, and~\ref{fig:params_other_models} represents a $y$-slice from the ensemble-averaged spatial profiles of any of the flow simulations tabulated in the SI, except for data points from homogeneous shear tests, which are the average of all data points in the simulation. To analyze simulations approximately at steady state, we only consider data satisfying $(\sigma_{xy}\dot\gamma+dQ_y/dy-\chi)/\chi<0.1$. To isolate heterogeneous flow and avoid blowup in the measurement $\kappa\equiv-\frac{Q_y}{dT/dy}$, only points with $|\alpha|>0.01$ are used to measure $\kappa$. As $\phi$ increases (a-d), diffusion becomes more prominent (larger $|\alpha|$). The critical packing fraction $\phi_0$ is postulated to be a material constant~\cite{berzi2024granular} that coincides on the $\mu(\phi)$ curve (Fig.~\ref{fig:scalings}d) with a critical value $\mu_0$. It is beyond this value that the $\phi$-based constitutive scalings most clearly lose accuracy that is recovered at least in part by the $(I,\Theta)$-based scalings.

The constitutive models shown as the gray curves in Fig.~\ref{fig:scalings}(d-g) are sufficient to close the system of equations for mass, momentum, and granular temperature (see SI Sec.~\ref{sec:model_predictions}~\cite{alessio2025supplemental} for example model solutions). Note the relation $\mu(\phi)$ in panel (d) is necessary to determine packing fraction, and the relation $\mu(I,\Theta)$ in panel (h) is redundant given the relation $\tilde\eta(I,\Theta)$ in panel (e). To calibrate this model, we perform three curve fits--$\mu(\phi)$, $\mu_0(I)$, and $\Theta_0(I)$--and determine the best-fitting exponents $a_{\eta}$, $a_{\chi}$, and $a_{\kappa}$ in the power law ansatzes, along with the constant parameter $A_\kappa$ that approximates $f_\kappa$. Any additional fitting would overdetermine the model. It is easy to show that $a_{\mu}=\frac12+a_\eta$, $\tilde\eta\Theta^{a_\eta}=\mu_0(I)\Theta_0(I)^{a_\mu}/I$, and $\tilde\chi\Theta^{a_\chi}=\mu_0(I)\Theta_0(I)^{a_\chi-\frac12}I$. The agreement between these expressions, plotted as gray lines in Fig.~\ref{fig:scalings}, and the DEM data indicates the model parametrization is robust. See SI~\cite{alessio2025supplemental} for details of the curve fittings and exponent estimations.

We show that a linearized form of Eq.~\ref{eqn:energy} recovers the NGF equation
\begin{equation}
    \xi^2\frac{d^2g}{dy^2} = g - g_0.
\end{equation}
Restricting our analysis to unidirectional flow varying along the spanwise direction $y$, Eq.~\ref{eqn:energy} can be expressed as 
\begin{equation}
\sqrt{\frac {p^3}{\rho d^2}}(\mu I - \tilde\chi\sqrt\Theta) + \frac{d}{dy}\left(\sqrt{\frac{p^3d^2}{\rho\Theta}}\tilde\kappa\frac{d\Theta}{dy}\right) = 0.
\end{equation}
Defining the local projection of the fluidity field $g_0$, we first show the difference $g_0-g$ derives from the imbalance between production and dissipation $\mu I - \tilde\chi\sqrt\Theta$. The definition of $g_0$ requires care~\cite{bouzid2015non}, and we proceed with two that carry qualitatively different consequences: the original NGF definition $g_0=I_0\sqrt{p/\rho d^2}/\mu$~\cite{kamrin2012nonlocal,henann2013predictive}, where $I_0$ is the inverse argument of the local rheology $\mu_0$, given $\mu$, and the recent EKT definition $g_0=\dot\gamma p /\tau_0$~\cite{berzi2024granular}, where $\tau_0=\chi/\dot\gamma$ is the homogeneous value of the shear stress. First with the EKT definition, noting $g/g_0=I\mu/\tilde\chi\sqrt\Theta$, it follows that $\mu I - \tilde\chi\sqrt\Theta = \frac{\tilde\chi\sqrt\Theta}{g}(g_0-g)$ exactly, as shown before for a specific version of EKT~\cite{berzi2024granular} and now for a general case (including the possibility of enduring contacts). For the original definition, we linearize about two base states: $I\to0$ and $I=I_0$, corresponding to $\mu<\mu_s$ (and $g_0=0$) and $\mu>\mu_s$. Utilizing our constitutive assumptions we identify for both cases $\mu I - \tilde\chi\sqrt\Theta = C (g_0-g)$, where $C$ is a base-state-specific prefactor with units of time (see SI~\cite{alessio2025supplemental} for details). We also find that the diffusion term $\xi^2\frac{d^2g}{dy^2}$, for cooperativity length $\xi$, derives from an approximation of the term $\frac{d}{dy}\left(\sqrt{\frac{p^3d^2}{\rho\Theta}}\tilde\kappa\frac{d\Theta}{dy}\right)$, identifying $\frac{\xi^2}{d^2}=\frac1{C}\sqrt{\frac{\rho d^2}p}\mu\sqrt{\left(\frac{\mu_0}\mu\right)^{a_\mu}\Theta_0}(\log\Theta_0)^\prime\tilde\kappa$.

Computing the parameters of the fluidity models~\cite{kamrin2014nonlocal} and heat equation~\cite{irmer2025granular} from our DEM data, we find disagreement with their forms that had been hypothesized in the literature (Fig.~\ref{fig:params_other_models}). The dependence of $\xi$ on the chosen definition of $g_0$ is consistent with our DEM data. Notating $\xi_\text{NGF}$ for the NGF definition~\cite{kamrin2012nonlocal} and $\xi_\text{EKT}$ for the EKT definition~\cite{berzi2024granular}, Fig.~\ref{fig:params_other_models}(a,b) plots our DEM measurements of $\xi$ vs. $\mu$, which are conditioned by the same criteria as our measurements of $\kappa$. (For this measurement we use the full, rather than linearized, form~\cite{kamrin2014nonlocal} of the NGF equation.) The NGF version (Fig.~\ref{fig:params_other_models}a) is singular at $\mu_s$ and the EKT version (Fig.~\ref{fig:params_other_models}b) is not, as evident also in the results of their references. Strikingly, both versions show significant spread from their previously hypothesized functional forms plotted as gray lines $\xi_\text{NGF}=A_\text{NGF}d\sqrt{\frac{\mu_2-\mu}{(\mu_2-\mu_s)|\mu_s-\mu|}}$~\cite{kamrin2014nonlocal}, for nonlocal amplitude $A_\text{NGF}$, and $\xi_\text{EKT}\approx 2d$~\cite{berzi2024granular}. In both cases the spread in the data is consistent with our derived expressions for $\xi$, which do not reduce to either of those simple forms. We demonstrate this further in Fig.~\ref{fig:params_other_models}(c,d). Within our data (Fig.~\ref{fig:params_other_models}c) we do not find $A_\text{NGF}$ to be a material constant because its value spans over a decade. Furthermore, ref.~\cite{berzi2024granular} reasoned based on the scarce measurements~\cite{oyama2019avalanche} of correlation length $\ell$ in the literature that $\ell/d$ may be nearly constant beyond the jamming fraction, concluding then that $\xi_\text{EKT}$ is approximately constant. However, our DEM data (Fig.~\ref{fig:params_other_models}d) shows that $\ell/d$ diverges beyond the jamming fraction, so further developments of EKT will require a careful investigation of the correlation length. We furthermore compute the parameters $D_\text{HE}$ and $B_\text{HE}$ of Ref.~\cite{irmer2025granular}'s heat equation, plotted in Fig.~\ref{fig:params_other_models}(e,f), finding considerable disagreement that would influence continuum models~\cite{yuan2026toward}. See SI~\cite{alessio2025supplemental} for more discussion and comparison of model solutions to DEM.

We finally remark that, like any constitutive model for granular flow, ours should not be considered universal even in the limited sense of steady unidirectional flow of stiff monodisperse spheres. Still, our DEM-validated closure relations prove consistent with popular fluidity models and constitute a necessary step toward a micromechanistic understanding of dense granular flows. In this framework, transport equations are developed by fitting DEM measurements of each term, taking full advantage of the DEM simulation, rather than by tuning phenomenological equations to predict fields observed in DEM. This procedure can be straightforwardly applied to characterize boundary conditions~\cite{jenkins1997flux,jing2016characterization}, a weakness in the current understanding that could be as consequential as the material rheology~\cite{debnath2022different,debnath2022comparison}. Where kinetic theory becomes intractable, our proposed framework can be applied to materials with complications such as polydispersity and nonsphericity~\cite{han2025rheology} for which future work will determine generalized transport equations. Generalizations of the closures accounting for unsteady fabric and dilation~\cite{sun2011constitutive,dsouza2020non,wiebicke2024simple,rojas2025transient,zhang2025micromechanics}, the possible role of fluctuating rotational motions~\cite{artoni2019coarse}, and external agitation~\cite{plati2026rheology}, remain prerequisite to broad application.

\begin{acknowledgments}
B.M.A. acknowledges financial support from the National Science Foundation Graduate Research Fellowship and the Stanford Graduate Fellowship. C.Y.L. was supported by the Alfred P. Sloan Foundation under grant number FG-2024-21649. B.M.A. and C.Y.L. acknowledge the Stanford Doerr School of Sustainability and the National Science Foundation Delta allocation MCH250068 computational resources and support. The authors thank Q. Zhang for helpful feedback.
\end{acknowledgments}

\bibliography{main}

@misc{itascaPFC,
  title={{PFC-Particle Flow Code, Ver. 7.0}},
  author={{Itasca Consulting Group, Inc.}},
  year={2021},
  note={{Minneapolis: Itasca}},
}

@article{hibler1977viscous,
  title={A viscous sea ice law as a stochastic average of plasticity},
  author={Hibler III, W. D.},
  journal={Journal of Geophysical Research},
  volume={82},
  number={27},
  pages={3932--3938},
  year={1977},
  publisher={Wiley Online Library}
}

@article{cundall1979discrete,
  title={A discrete numerical model for granular assemblies},
  author={Cundall, Peter A and Strack, Otto D. L.},
  journal={Geotechnique},
  volume={29},
  number={1},
  pages={47--65},
  year={1979},
  publisher={Thomas Telford Ltd}
}

@article{johnson1987frictional,
  title={Frictional--collisional constitutive relations for granular materials, with application to plane shearing},
  author={Johnson, Paul C and Jackson, Roy},
  journal={Journal of {F}luid {M}echanics},
  volume={176},
  pages={67--93},
  year={1987},
  publisher={Cambridge University Press}
}

@article{tsuji1992lagrangian,
  title={Lagrangian numerical simulation of plug flow of cohesionless particles in a horizontal pipe},
  author={Tsuji, Yutaka and Tanaka, Toshitsugu and Ishida, T},
  journal={Powder Technology},
  volume={71},
  number={3},
  pages={239--250},
  year={1992},
  publisher={Elsevier}
}

@article{savage1992instability,
  title={Instability of unbounded uniform granular shear flow},
  author={Savage, S. B.},
  journal={Journal of Fluid Mechanics},
  volume={241},
  pages={109--123},
  year={1992},
  publisher={Cambridge University Press}
}

@article{jenkins1997flux,
  title={On the flux of fluctuation energy in a collisional grain flow at a flat, frictional wall},
  author={Jenkins, James T and Louge, Michel Y},
  journal={Physics of Fluids},
  volume={9},
  number={10},
  pages={2835--2840},
  year={1997},
  publisher={American Institute of Physics}
}

@article{savage1998analyses,
  title={Analyses of slow high-concentration flows of granular materials},
  author={Savage, S. B.},
  journal={Journal of Fluid Mechanics},
  volume={377},
  pages={1--26},
  year={1998},
  publisher={Cambridge University Press}
}

@article{wolf1998dissipation,
  title={Dissipation in granular materials},
  author={Wolf, Dietrich E and Radjai, Farhang and Dippel, Sabine},
  journal={Philosophical Magazine B},
  volume={77},
  number={5},
  pages={1413--1425},
  year={1998},
  publisher={Taylor \& Francis}
}

@article{garzo1999dense,
  title={Dense fluid transport for inelastic hard spheres},
  author={Garz{\'o}, V and Dufty, J. W.},
  journal={Physical Review E},
  volume={59},
  number={5},
  pages={5895},
  year={1999},
  publisher={APS}
}

@article{barber2000contact,
  title={Contact mechanics},
  author={Barber, J. R. and Ciavarella, M},
  journal={International Journal of Solids and Structures},
  volume={37},
  number={1-2},
  pages={29--43},
  year={2000},
  publisher={Elsevier}
}

@article{losert2000particle,
  title={Particle dynamics in sheared granular matter},
  author={Losert, Wolfgang and Bocquet, Lyderic and Lubensky, Tom C and Gollub, Jerry P},
  journal={Physical Review Letters},
  volume={85},
  number={7},
  pages={1428},
  year={2000},
  publisher={APS}
}

@article{bocquet2001granular,
  title={Granular shear flow dynamics and forces: Experiment and continuum theory},
  author={Bocquet, Lyderic and Losert, Wolfgang and Schalk, David and Lubensky, T. C. and Gollub, J. P.},
  journal={Physical Review E},
  volume={65},
  number={1},
  pages={011307},
  year={2001},
  publisher={APS}
}

@article{pope2001turbulent,
  title={Turbulent flows},
  author={Pope, Stephen B},
  journal={Measurement Science and Technology},
  volume={12},
  number={11},
  pages={2020--2021},
  year={2001}
}

@article{campbell2002granular,
  title={Granular shear flows at the elastic limit},
  author={Campbell, Charles S},
  journal={Journal of Fluid Mechanics},
  volume={465},
  pages={261--291},
  year={2002},
  publisher={Cambridge University Press}
}

@article{radjai2002turbulentlike,
  title={Turbulentlike fluctuations in quasistatic flow of granular media},
  author={Radjai, Farhang and Roux, St{\'e}phane},
  journal={Physical Review Letters},
  volume={89},
  number={6},
  pages={064302},
  year={2002},
  publisher={APS}
}

@article{aranson2002continuum,
  title={Continuum theory of partially fluidized granular flows},
  author={Aranson, Igor S and Tsimring, Lev S},
  journal={Physical Review E},
  volume={65},
  number={6},
  pages={061303},
  year={2002},
  publisher={APS}
}

@article{da2005rheophysics,
  title={Rheophysics of dense granular materials: Discrete simulation of plane shear flows},
  author={da Cruz, Fr{\'e}d{\'e}ric and Emam, Sacha and Prochnow, Micha{\"e}l and Roux, Jean-No{\"e}l and Chevoir, Fran{\c{c}}ois},
  journal={Physical Review E},
  volume={72},
  number={2},
  pages={021309},
  year={2005},
  publisher={APS}
}

@article{bazant2006spot,
  title={The spot model for random-packing dynamics},
  author={Bazant, Martin Z},
  journal={Mechanics of Materials},
  volume={38},
  number={8-10},
  pages={717--731},
  year={2006},
  publisher={Elsevier}
}

@article{jop2006constitutive,
  title={A constitutive law for dense granular flows},
  author={Jop, Pierre and Forterre, Yo{\"e}l and Pouliquen, Olivier},
  journal={Nature},
  volume={441},
  number={7094},
  pages={727--730},
  year={2006},
  publisher={Nature Publishing Group UK London}
}

@article{jenkins2006dense,
  title={Dense shearing flows of inelastic disks},
  author={Jenkins, James T},
  journal={Physics of Fluids},
  volume={18},
  number={10},
  year={2006},
  publisher={AIP Publishing}
}

@article{jenkins2007dense,
  title={Dense inclined flows of inelastic spheres},
  author={Jenkins, James T},
  journal={Granular Matter},
  volume={10},
  number={1},
  pages={47--52},
  year={2007},
  publisher={Springer}
}

@book{j2007statistical,
  title={Statistical mechanics of nonequilibrium liquids},
  author={Evans, Denis J. and Morriss, Gary P.},
  year={2007},
  publisher={ANU Press}
}

@article{goyon2008spatial,
  title={Spatial cooperativity in soft glassy flows},
  author={Goyon, Julie and Colin, Annie and Ovarlez, G and Ajdari, A and Bocquet, L},
  journal={Nature},
  volume={454},
  number={7200},
  pages={84--87},
  year={2008},
  publisher={Nature Publishing Group UK London}
}

@article{luding2008introduction,
  title={Introduction to discrete element methods: basic of contact force models and how to perform the micro-macro transition to continuum theory},
  author={Luding, Stefan},
  journal={European Journal of Environmental and Civil Engineering},
  volume={12},
  number={7-8},
  pages={785--826},
  year={2008},
  publisher={Taylor \& Francis}
}

@article{bocquet2009kinetic,
  title={Kinetic theory of plastic flow in soft glassy materials},
  author={Bocquet, Lyd{\'e}ric and Colin, Annie and Ajdari, Armand},
  journal={Physical Review Letters},
  volume={103},
  number={3},
  pages={036001},
  year={2009},
  publisher={APS}
}

@article{pouliquen2009non,
  title={A non-local rheology for dense granular flows},
  author={Pouliquen, Olivier and Forterre, Yoel},
  journal={Philosophical Transactions of the Royal Society A: Mathematical, Physical and Engineering Sciences},
  volume={367},
  number={1909},
  pages={5091--5107},
  year={2009},
  publisher={The Royal Society Publishing}
}

@article{goldhirsch2010stress,
  title={Stress, stress asymmetry and couple stress: from discrete particles to continuous fields},
  author={Goldhirsch, Isaac},
  journal={Granular Matter},
  volume={12},
  number={3},
  pages={239--252},
  year={2010},
  publisher={Springer}
}

@book{brilliantov2010kinetic,
  title={Kinetic theory of granular gases},
  author={Brilliantov, Nikolai V and P{\"o}schel, Thorsten},
  year={2010},
  publisher={Oxford University Press, USA}
}

@article{zhang2010coarse,
  title={Coarse-graining of a physical granular system},
  author={Zhang, Jie and Behringer, Robert P and Goldhirsch, Isaac},
  journal={Progress of Theoretical Physics Supplement},
  volume={184},
  pages={16--30},
  year={2010},
  publisher={Oxford University Press}
}

@article{sun2011constitutive,
  title={A constitutive model with microstructure evolution for flow of rate-independent granular materials},
  author={Sun, Jin and Sundaresan, Sankaran},
  journal={Journal of Fluid Mechanics},
  volume={682},
  pages={590--616},
  year={2011},
  publisher={Cambridge University Press}
}

@article{berzi2011surface,
  title={Surface flows of inelastic spheres},
  author={Berzi, Diego and Jenkins, James T},
  journal={Physics of Fluids},
  volume={23},
  number={1},
  year={2011},
  publisher={AIP Publishing}
}

@article{berzi2011constitutive,
  title={Constitutive relations for steady, dense granular flows},
  author={Berzi, Diego and di Prisco, Claudio Giulio and Vescovi, Dalila},
  journal={Physical Review E},
  volume={84},
  number={3},
  pages={031301},
  year={2011},
  publisher={APS}
}

@article{dijksman2011jamming,
  title={Jamming, yielding, and rheology of weakly vibrated granular media},
  author={Dijksman, Joshua A and Wortel, Geert H and van Dellen, Louwrens T. H. and Dauchot, Olivier and van Hecke, Martin},
  journal={Physical Review Letters},
  volume={107},
  number={10},
  pages={108303},
  year={2011},
  publisher={APS}
}

@article{chialvo2012bridging,
  title={Bridging the rheology of granular flows in three regimes},
  author={Chialvo, Sebastian and Sun, Jin and Sundaresan, Sankaran},
  journal={Physical Review E},
  volume={85},
  number={2},
  pages={021305},
  year={2012},
  publisher={APS}
}

@article{kamrin2012nonlocal,
  title={Nonlocal constitutive relation for steady granular flow},
  author={Kamrin, Ken and Koval, Georg},
  journal={Physical Review Letters},
  volume={108},
  number={17},
  pages={178301},
  year={2012},
  publisher={APS}
}

@article{weinhart2013coarse,
  title={Coarse-grained local and objective continuum description of three-dimensional granular flows down an inclined surface},
  author={Weinhart, Thomas and Hartkamp, Remco and Thornton, Anthony R and Luding, Stefan},
  journal={Physics of Fluids},
  volume={25},
  number={7},
  year={2013},
  publisher={AIP Publishing}
}

@article{bouzid2013nonlocal,
  title={Nonlocal rheology of granular flows across yield conditions},
  author={Bouzid, Mehdi and Trulsson, Martin and Claudin, Philippe and Cl{\'e}ment, Eric and Andreotti, Bruno},
  journal={Physical Review Letters},
  volume={111},
  number={23},
  pages={238301},
  year={2013},
  publisher={APS}
}

@article{henann2013predictive,
  title={A predictive, size-dependent continuum model for dense granular flows},
  author={Henann, David L and Kamrin, Ken},
  journal={Proceedings of the National Academy of Sciences},
  volume={110},
  number={17},
  pages={6730--6735},
  year={2013},
  publisher={National Academy of Sciences}
}

@article{miller2013eddy,
  title={Eddy viscosity in dense granular flows},
  author={Miller, Thomas and Rognon, P and Metzger, B and Einav, I},
  journal={Physical Review Letters},
  volume={111},
  number={5},
  pages={058002},
  year={2013},
  publisher={APS}
}

@article{chialvo2013modified,
  title={A modified kinetic theory for frictional granular flows in dense and dilute regimes},
  author={Chialvo, Sebastian and Sundaresan, Sankaran},
  journal={Physics of Fluids},
  volume={25},
  number={7},
  year={2013},
  publisher={AIP Publishing}
}

@article{kamrin2014nonlocal,
  title={Nonlocal modeling of granular flows down inclines},
  author={Kamrin, Ken and Henann, David L},
  journal={Soft Matter},
  volume={11},
  number={1},
  pages={179--185},
  year={2014},
  publisher={The Royal Society of Chemistry}
}

@article{bouzid2015non,
  title={Non-local rheology in dense granular flows: Revisiting the concept of fluidity},
  author={Bouzid, Mehdi and Izzet, Adrien and Trulsson, Martin and Cl{\'e}ment, Eric and Claudin, Philippe and Andreotti, Bruno},
  journal={The European Physical Journal E},
  volume={38},
  pages={1--15},
  year={2015},
  publisher={Springer}
}

@article{rognon2015long,
  title={Long-range wall perturbations in dense granular flows},
  author={Rognon, Pierre G and Miller, Thomas and Metzger, Bloen and Einav, Itai},
  journal={Journal of Fluid Mechanics},
  volume={764},
  pages={171--192},
  year={2015},
  publisher={Cambridge University Press}
}

@article{berzi2015steady,
  title={Steady shearing flows of deformable, inelastic spheres},
  author={Berzi, Diego and Jenkins, James T},
  journal={Soft Matter},
  volume={11},
  number={24},
  pages={4799--4808},
  year={2015},
  publisher={Royal Society of Chemistry}
}

@article{berzi2015different,
  title={Different singularities in the functions of extended kinetic theory at the origin of the yield stress in granular flows},
  author={Berzi, Diego and Vescovi, Dalila},
  journal={Physics of {F}luids},
  volume={27},
  number={1},
  year={2015},
  publisher={AIP Publishing}
}

@article{pahtz2015fluctuation,
  title={The fluctuation energy balance in non-suspended fluid-mediated particle transport},
  author={P{\"a}htz, Thomas and Dur{\'a}n, Orencio and Ho, Tuan-Duc and Valance, Alexandre and Kok, Jasper F},
  journal={Physics of Fluids},
  volume={27},
  number={1},
  year={2015},
  publisher={AIP Publishing}
}

@article{jing2016characterization,
  title={Characterization of base roughness for granular chute flows},
  author={Jing, Lu and Kwok, C. Y. and Leung, Yat Fai and Sobral, Y. D.},
  journal={Physical Review E},
  volume={94},
  number={5},
  pages={052901},
  year={2016},
  publisher={APS}
}

@article{saitoh2016anomalous,
  title={Anomalous energy cascades in dense granular materials yielding under simple shear deformations},
  author={Saitoh, Kuniyasu and Mizuno, Hideyuki},
  journal={Soft Matter},
  volume={12},
  number={5},
  pages={1360--1367},
  year={2016},
  publisher={Royal Society of Chemistry}
}

@article{vescovi2016merging,
  title={Merging fluid and solid granular behavior},
  author={Vescovi, Dalila and Luding, Stefan},
  journal={Soft Matter},
  volume={12},
  number={41},
  pages={8616--8628},
  year={2016},
  publisher={Royal Society of Chemistry}
}

@article{wortel2016criticality,
  title={Criticality in vibrated frictional flows at a finite strain rate},
  author={Wortel, Geert and Dauchot, Olivier and van Hecke, Martin},
  journal={Physical Review Letters},
  volume={117},
  number={19},
  pages={198002},
  year={2016},
  publisher={APS}
}

@article{govender2016granular,
  title={Granular flows in rotating drums: A rheological perspective},
  author={Govender, Indresan},
  journal={Minerals Engineering},
  volume={92},
  pages={168--175},
  year={2016},
  publisher={Elsevier}
}

@article{degiuli2017friction,
  title={Friction law and hysteresis in granular materials},
  author={DeGiuli, E and Wyart, M},
  journal={Proceedings of the National Academy of Sciences},
  volume={114},
  number={35},
  pages={9284--9289},
  year={2017},
  publisher={National Academy of Sciences}
}

@article{saitoh2017anisotropic,
  title={Anisotropic decay of the energy spectrum in two-dimensional dense granular flows},
  author={Saitoh, Kuniyasu and Mizuno, Hideyuki},
  journal={Physical Review E},
  volume={96},
  number={1},
  pages={012903},
  year={2017},
  publisher={APS}
}

@article{zhang2017microscopic,
  title={Microscopic description of the granular fluidity field in nonlocal flow modeling},
  author={Zhang, Qiong and Kamrin, Ken},
  journal={Physical Review Letters},
  volume={118},
  number={5},
  pages={058001},
  year={2017},
  publisher={APS}
}

@article{berzi2018fluidity,
  title={Fluidity, anisotropy, and velocity correlations in frictionless, collisional grain flows},
  author={Berzi, Diego and Jenkins, James T},
  journal={Physical Review Fluids},
  volume={3},
  number={9},
  pages={094303},
  year={2018},
  publisher={APS}
}

@article{artoni2019coarse,
  title={Coarse graining for granular materials: micro-polar balances},
  author={Artoni, Riccardo and Richard, Patrick},
  journal={Acta Mechanica},
  volume={230},
  number={9},
  pages={3055--3069},
  year={2019},
  publisher={Springer}
}

@article{pahtz2019local,
  title={Local rheology relation with variable yield stress ratio across dry, wet, dense, and dilute granular flows},
  author={P{\"a}htz, Thomas and Dur{\'a}n, Orencio and De Klerk, David N and Govender, Indresan and Trulsson, Martin},
  journal={Physical Review Letters},
  volume={123},
  number={4},
  pages={048001},
  year={2019},
  publisher={APS}
}

@article{berzi2019erodible,
  title={Erodible, granular beds are fragile},
  author={Berzi, Diego and Jenkins, James T and Richard, Patrick},
  journal={Soft Matter},
  volume={15},
  number={36},
  pages={7173--7178},
  year={2019},
  publisher={Royal Society of Chemistry}
}

@article{lorenz2019theory,
  title={A theory of angel hair: Analytic prediction of frictional heating of particulates in pneumatic transport},
  author={Lorenz, Ralph D},
  journal={Powder Technology},
  volume={355},
  pages={264--267},
  year={2019},
  publisher={Elsevier}
}

@article{oyama2019avalanche,
  title={Avalanche interpretation of the power-law energy spectrum in three-dimensional dense granular flow},
  author={Oyama, Norihiro and Mizuno, Hideyuki and Saitoh, Kuniyasu},
  journal={Physical Review Letters},
  volume={122},
  number={18},
  pages={188004},
  year={2019},
  publisher={APS}
}

@article{duan2019new,
  title={A new kinetic theory model of granular flows that incorporates particle stiffness},
  author={Duan, Yifei and Feng, Zhi-Gang},
  journal={Physics of Fluids},
  volume={31},
  number={1},
  year={2019},
  publisher={AIP Publishing}
}

@article{jerolmack2019viewing,
  title={Viewing {Earth}’s surface as a soft-matter landscape},
  author={Jerolmack, Douglas J and Daniels, Karen E},
  journal={Nature Reviews Physics},
  volume={1},
  number={12},
  pages={716--730},
  year={2019},
  publisher={Nature Publishing Group UK London}
}

@article{hu2019taichi,
  title={{Taichi}: a language for high-performance computation on spatially sparse data structures},
  author={Hu, Yuanming and Li, Tzu-Mao and Anderson, Luke and Ragan-Kelley, Jonathan and Durand, Fr{\'e}do},
  journal={ACM Transactions on Graphics (TOG)},
  volume={38},
  number={6},
  pages={1--16},
  year={2019},
  publisher={ACM New York, NY, USA}
}

@article{dsouza2020non,
  title={A non-local constitutive model for slow granular flow that incorporates dilatancy},
  author={Dsouza, Peter Varun and Nott, Prabhu R},
  journal={Journal of Fluid Mechanics},
  volume={888},
  pages={R3},
  year={2020},
  publisher={Cambridge University Press}
}

@article{berzi2020extended,
  title={Extended kinetic theory for granular flow over and within an inclined erodible bed},
  author={Berzi, Diego and Jenkins, James T and Richard, Patrick},
  journal={Journal of Fluid Mechanics},
  volume={885},
  pages={A27},
  year={2020},
  publisher={Cambridge University Press}
}

@article{gaume2020microscopic,
  title={Microscopic origin of nonlocal rheology in dense granular materials},
  author={Gaume, Johan and Chambon, Guillaume and Naaim, Mohamed},
  journal={Physical Review Letters},
  volume={125},
  number={18},
  pages={188001},
  year={2020},
  publisher={APS}
}

@article{kim2020power,
  title={Power-law scaling in granular rheology across flow geometries},
  author={Kim, Seongmin and Kamrin, Ken},
  journal={Physical Review Letters},
  volume={125},
  number={8},
  pages={088002},
  year={2020},
  publisher={APS}
}

@article{robinson2021examination,
  title={Examination of the microscopic definition for granular fluidity},
  author={Robinson, James A and Holland, Daniel J and Fullard, Luke},
  journal={Physical Review Fluids},
  volume={6},
  number={4},
  pages={044302},
  year={2021},
  publisher={APS}
}

@article{fazelpour2022effect,
  title={The effect of grain shape and material on the nonlocal rheology of dense granular flows},
  author={Fazelpour, Farnaz and Tang, Zhu and Daniels, Karen E},
  journal={Soft Matter},
  volume={18},
  number={7},
  pages={1435--1442},
  year={2022},
  publisher={Royal Society of Chemistry}
}

@article{debnath2022different,
  title={Different shear regimes in the dense granular flow in a vertical channel},
  author={Debnath, Bhanjan and Rao, K Kesava and Kumaran, V},
  journal={Journal of Fluid Mechanics},
  volume={945},
  pages={A25},
  year={2022},
  publisher={Cambridge University Press}
}

@article{debnath2022comparison,
  title={Comparison of the compressible class of models and non-local models with the discrete element method for steady fully developed flow of cohesionless granular materials through a vertical channel},
  author={Debnath, Bhanjan and Kumaran, V and Rao, K Kesava},
  journal={Journal of Fluid Mechanics},
  volume={937},
  pages={A33},
  year={2022},
  publisher={Cambridge University Press}
}

@article{rangel2023experimental,
  title={Experimental investigation of heat generation during granular flow in a rotating drum using infrared thermography},
  author={Rangel, Rafael L and Kisuka, Francisco and Hare, Colin and Vivacqua, Vincenzino and Franci, Alessandro and O{\~n}ate, Eugenio and Wu, Chuan-Yu},
  journal={Powder Technology},
  volume={426},
  pages={118619},
  year={2023},
  publisher={Elsevier}
}

@article{zaccone2023general,
  title={General theory of the viscosity of liquids and solids from nonaffine particle motions},
  author={Zaccone, Alessio},
  journal={Physical Review E},
  volume={108},
  pages={044101},
  year={2023},
  publisher={APS}
}

@article{poon2023microscopic,
  title={Microscopic origin of granular fluidity: An experimental investigation},
  author={Poon, Rebecca N and Thomas, Amalia L and Vriend, Nathalie M},
  journal={Physical Review E},
  volume={108},
  number={6},
  pages={064902},
  year={2023},
  publisher={APS}
}

@article{chassagne2023frictional,
  title={A frictional--collisional model for bedload transport based on kinetic theory of granular flows: discrete and continuum approaches},
  author={Chassagne, R{\'e}mi and Bonamy, Cyrille and Chauchat, Julien},
  journal={Journal of Fluid Mechanics},
  volume={964},
  pages={A27},
  year={2023},
  publisher={Cambridge University Press}
}

@article{wiebicke2024simple,
  title={A simple hydrodynamic model for clay},
  author={Wiebicke, Max and Einav, Itai},
  journal={Journal of the Mechanics and Physics of Solids},
  volume={192},
  pages={105789},
  year={2024},
  publisher={Elsevier}
}

@article{kamrin2024advances,
  title={Advances in Modeling Dense Granular Media},
  author={Kamrin, Ken and Hill, Kimberly M and Goldman, Daniel I and Andrade, Jose E},
  journal={Annual Review of Fluid Mechanics},
  volume={56},
  pages={215--240},
  year={2024},
  publisher={Annual Reviews}
}

@article{berzi2024granular,
  title={On granular flows: From kinetic theory to inertial rheology and nonlocal constitutive models},
  author={Berzi, Diego},
  journal={Physical Review Fluids},
  volume={9},
  number={3},
  pages={034304},
  year={2024},
  publisher={APS}
}

@article{vescovi2024extended,
  title={Extended kinetic theory applied to pressure-controlled shear flows of frictionless spheres between rigid, bumpy planes},
  author={Vescovi, Dalila and de Wijn, Astrid S and Cross, Graham L. W. and Berzi, Diego},
  journal={Soft Matter},
  volume={20},
  number={43},
  pages={8702--8715},
  year={2024},
  publisher={Royal Society of Chemistry}
}

@article{irmer2025granular,
  title={Granular temperature controls local rheology of vibrated granular flows},
  author={Irmer, Mitchell G and Brodsky, Emily E and Clark, Abram H},
  journal={Physical Review Letters},
  volume={134},
  number={4},
  pages={048202},
  year={2025},
  publisher={APS}
}

@article{gautam2025micromechanical,
  title={Micromechanical Origin of Rate Independence of the Stress in Sheared Granular Materials},
  author={Gautam, Ravi and Nott, Prabhu R},
  journal={Physical Review Letters},
  volume={134},
  number={15},
  pages={158201},
  year={2025},
  publisher={APS}
}

@article{rojas2025transient,
  title={Transient stress and fabric model for quasi-static granular flows in three dimensions},
  author={Rojas, Eduardo and Kamrin, Ken},
  journal={Soft Matter},
  volume={21},
  number={15},
  pages={2896--2908},
  year={2025},
  publisher={Royal Society of Chemistry}
}

@article{zhang2025micromechanics,
  title={Micromechanics-inspired granular thermodynamics: A constitutive model for multidirectional cyclic shearing},
  author={Zhang, Zhichao and Soga, Kenichi},
  journal={Journal of the Mechanics and Physics of Solids},
  pages={106170},
  year={2025},
  publisher={Elsevier}
}

@article{han2025rheology,
  title={Rheology of nonconvex granular flows based on particle rotational characteristics},
  author={Han, Wenjin and Zhao, He and Wang, Dengming},
  journal={Physical Review E},
  volume={111},
  number={1},
  pages={015415},
  year={2025},
  publisher={APS}
}

@misc{alessio2025supplemental,
  title={Supplemental Material for ``{Dense} granular rheology from fluctuations''},
  author={Alessio, Benjamin M. and Edwards, Matthew R. and Lai, Ching-Yao},
  note={See attached Supplemental Material},
  year={2026}
}

@article{yuan2026toward,
  title={Toward a unified continuum framework for dense granular flows: rheology extended with granular temperature},
  author={Yuan, Zexu and Zhao, He and Wang, Dengming},
  journal={Journal of Fluid Mechanics},
  volume={1030},
  pages={A24},
  year={2026},
  publisher={Cambridge University Press}
}

@article{plati2026rheology,
  title={Rheology of dense vibrated granular flows: Nonmonotonic response controlled by granular temperature},
  author={Plati, Andrea and Petrillo, Giuseppe and de Arcangelis, Lucilla and Gnoli, Andrea and Puglisi, Andrea and Sarracino, Alessandro and Lippiello, E},
  journal={Physical Review Research},
  volume={8},
  number={3},
  pages={L032010},
  year={2026},
  publisher={APS}
}

\clearpage

\appendix
\onecolumngrid

\begin{center}
\huge Supplemental Material
\end{center}

\section{Simulations}

\subsection{Discrete element methodology}
\label{sub:discrete_element_methodology}
We use discrete element method (DEM) simulations~\cite{luding2008introduction} in three dimensions to model spherical grains, implemented in house using Taichi~\cite{hu2019taichi} drawing from their minimal DEM example. We implement the Hertz-Mindlin model~\cite{barber2000contact} for elastic and frictional contact forces and the Tsuji model~\cite{tsuji1992lagrangian} for dashpot forces. The forces and resulting trajectories are computed for each grain using the following summarized formulation. 

A pair of contacting, spherical grains have radii $r_1$ and $r_2$, masses $m_1$ and $m_2$, effective mass $m=m_1m_2/(m_1+m_2)$, effective radius $r=r_1r_2/(r_1+r_2)$, center locations $\boldsymbol x_1$ and $\boldsymbol x_2$, center-to-contact separation vectors $\boldsymbol\ell_1$ and $\boldsymbol \ell_2$, contact relative velocity vector $\boldsymbol v$ (including both translation and rotation $\boldsymbol\omega_1\times\boldsymbol \ell_1+\boldsymbol\omega_2\times\boldsymbol\ell_2$), and overlap distance $\boldsymbol\delta$. The unit vector specifying the direction normal to the contact surface is $\boldsymbol e_{\parallel}$ so that the normal component of the velocity is $\boldsymbol v_{\parallel}=(\boldsymbol v\boldsymbol\cdot\boldsymbol e_{\parallel})\boldsymbol e_{\parallel}$ and the tangential component is $\boldsymbol v_\perp=\boldsymbol v - \boldsymbol v_{\parallel}$. Each grain is given the same Young's modulus $E_0$ and Poisson's ratio $\nu$, so that the effective Young's modulus is $E=E_0/2(1-\nu^2)$, and the effective spring stiffness is $k_\parallel=\frac43r\sqrt{\delta_\parallel }$. Given coefficient of restitution $e$, parameter $A = 1.2728 - 4.2783e + 11.087e^2 - 22.348e^3 + 27.467e^4 - 18.022e^5 + 4.8218e^6$, and normal damping coefficient $\gamma_{\parallel}=A\sqrt{m\frac43E\sqrt{r\delta}}$, the force $\boldsymbol F_n$ directed normal to the contact surface is $\boldsymbol F_n = k_\parallel\delta_\parallel \boldsymbol e_{\parallel} - \gamma_{\parallel}\boldsymbol v_{\parallel}$, where we identify the first and second terms on the right hand side as the elastic and dissipative parts, respectively. A limit on the normal dissipative force is imposed to prevent intergrain attraction. We also track the tangential overlap $\delta_\perp $ at each time step by accumulating as $\boldsymbol\delta_\perp (t) = \boldsymbol\delta_\perp (t-\Delta t) + \boldsymbol v_\perp(t-\Delta t)\Delta t$, noting that $\boldsymbol\delta_\perp $ is a vector quantity because we must keep track of its orientation on the tangential plane. After incrementing, we also remove the normal component while preserving the magnitude by subtracting $\boldsymbol\delta_\perp (t)\boldsymbol\cdot\boldsymbol e_{\parallel}$ and multiplying by $||\boldsymbol\delta_\perp (t-\Delta t)||/||\boldsymbol\delta_\perp (t)||$. The tangential stiffness $k_\perp $ and damping coefficient $\gamma_\perp $ are computed in several substeps. First, we assign $k_\perp =2E\sqrt{r\delta_\parallel }/((1+\nu)(2-\nu))$ and, for tangential damping ratio $0<A_t<1$, $\gamma_\perp =A_t\gamma_{\parallel}$. Then, for $B=\min(1, \sqrt{\delta_\parallel (t)/\delta_\parallel (t-\Delta t)})$, we compute the elastic tangential force $\boldsymbol F_{te}$ directly as $\boldsymbol F_{te}(t)=B\boldsymbol F_{te}(t-\Delta t)-k_\perp (\boldsymbol \delta_\perp (t)-\boldsymbol \delta_\perp (t-\Delta t))$. For vector norm $||\boldsymbol f||=\sqrt{\boldsymbol f\boldsymbol\cdot\boldsymbol f}$, we recalculate $k_\perp =||\boldsymbol F_{te}||/||\boldsymbol\delta_\perp ||$. Then, a trial force is computed as $\boldsymbol F_t = k_\perp \boldsymbol\delta_\perp  - \gamma_\perp \boldsymbol v_\perp$. Finally, we apply a frictional truncation for grain-grain friction coefficient $\mu_g$. If $||\boldsymbol F_{te}||>\mu_g||\boldsymbol F_n||$, then we multiply $\gamma_\perp $ and $\boldsymbol\delta_\perp $ by $\mu_g||\boldsymbol F_n||/||\boldsymbol F_{te}||$ and recompute $\boldsymbol F_t = k_\perp \boldsymbol\delta_\perp  - \gamma_\perp \boldsymbol v_\perp$. The total force $\boldsymbol F=\boldsymbol F_n+\boldsymbol F_t$ is equal and opposite for both grains. The torques are computed as $\boldsymbol\ell_1\times \boldsymbol F$ and $-\boldsymbol\ell_2\times\boldsymbol F$. For a given configuration, once the forces and torques are calculated (also including gravitational body force $m_g\boldsymbol G$), a velocity-Verlet algorithm is used to compute translational and angular velocities and advance to the next time step.

Bounding sliding walls are implemented as rotationless rigid clumps. The velocity components are set either as a sliding speed (in the streamwise direction) or as a mechanism to control pressure (in the dilating direction $y$) using servo control~\cite{da2005rheophysics} $U_y=-(F_{y,top}-F_\text{desired})/\gamma_{sc}$.

We compute the rate of dissipation of mechanical energy per contact $W_c$ as the sum of two contributions. The first is that from the viscous damping $W_c^\text{dash}=\gamma_{\parallel}||\boldsymbol v_{\parallel}||^2 + \gamma_\perp ||\boldsymbol v_\perp||^2$. The second is from the frictional slip, at speed $\boldsymbol w=\left((\boldsymbol v\boldsymbol \cdot\boldsymbol e_\perp)\boldsymbol e_{\perp} \Delta t - (\boldsymbol \delta_\perp (t)-\boldsymbol\delta_\perp (t-\Delta t))\right)/\Delta t$, expressed as $W_c^\text{fric}=\frac12\boldsymbol w\boldsymbol\cdot\left(\boldsymbol F_{te}(t)+\boldsymbol F_{te}(t-\Delta t)\right)$~\cite{itascaPFC}. We also compute per contact the strain energy from normal $E_c^\parallel=\frac12k_\parallel||\delta_\parallel||^2$ and tangential $E_c^\perp = \frac12k_\perp ||\delta_\perp ||^2$ compression, and per grain the gravitational potential energy $E_g^\text{grav} = -m_g\boldsymbol G\boldsymbol\cdot\boldsymbol x_g$, translational kinetic energy $E_g^\text{tke}=\frac12m_g||\boldsymbol v_g||^2$, and rotational kinetic energy $E_g^\text{rke}=\frac15m_gr_g^2||\boldsymbol \omega_g||^2$.

\subsection{Material parameters}
For our simulated material, with arbitrary units for all dimensional quantities we choose coefficient of restitution $e=0.7$, tangential damping ratio $A_t=0.25$, mean diameter $d=1$ with uniform polydispersity of $\pm$15\%, density $\rho=6/\pi$, friction coefficient $\mu_g=0.15$, effective Young's modulus $E=1$, Poisson's ratio $\nu=0.3$, and servo control damping $\gamma_{sc}=10$.

\begin{figure}
\includegraphics[width=0.99\textwidth]{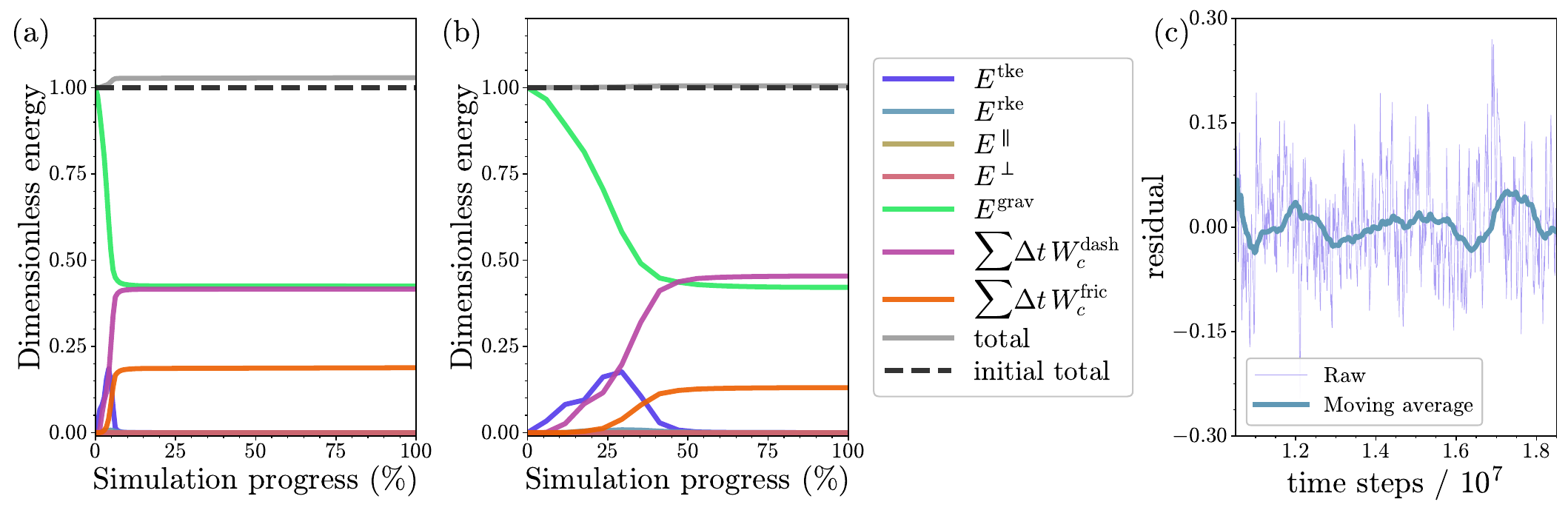}
\caption{\label{fig:calibration}Benchmarking the convergence of total energy error for decreasing time step. (a)~$\Delta t=0.8$ has a long-time error of 2.7\% and (b)~$\Delta t=0.2$ of 0.5\%. Each term is nondimensionalized by the initial total energy, and represents the sum across all grains or contacts. (c) Example time series of the energy residual, including a moving average over a window of 100 time steps. A time average of zero indicates statistical stationarity. }
\end{figure}

To determine an appropriate time step, we construct a system of 200 grains periodic in $x$ and $z$, initially not touching, settling under gravity $G=10^{-7}$ in $-y$ bounded by an elastically bouncy floor. The initial velocities are drawn randomly for each component from a uniform distribution with mean 0 and standard deviation $10^{-4}$. The total energy, including translational kinetic, rotational kinetic, gravitational potential, normal potential, tangential potential, dashpot-dissipated, and friction-dissipated, is computed by summing across all grains $E=\sum_gE_g$ or contacts $E=\sum_cE_c$, and tracked over time. Figure~\ref{fig:calibration} shows the time series for two different choices of time step. We select $\Delta t=0.2$, considering 0.5\% to be an acceptable error.

\subsection{Parameter sweep}

Large randomly packed piles are generated by gravitational settling in semiperiodic domains. Subdomains of these piles are exported as shear cells, where top and bottom plates composed of frozen-together slices of the packing sandwich a granular medium. The spatial dimension of the shear cells is fixed in the periodic directions $x$ and $z$ with length $L_x=L_z=25d$. The height $L_y$ is determined by the number of grains $N$, and to a lesser extent by packing fraction which is affected by the flow. The flow is driven by the bottom plate sliding in $-x$ at speed $U$, and an applied acceleration (i.e. gravity) of magnitude $G$ and direction $\psi$ (degrees counterclockwise from $-y$). Pressure $P$ is imposed, as explained in Sec.~\ref{sub:discrete_element_methodology}, by servo control with $F_\text{desired}=PL_xL_z$. This sets the boundary condition at the plate for the pressure field $p$, which is also impacted by the $-y$ component of gravity. In some cases, there is a spatially varying body force (as introduced by Ref.~\cite{kim2020power}), for midpoint $y_m$ as $f_b=G(y-y_m)/L_y$, with care taken not to overcome the servo pressure. These were included because they contribute many points where production outpaces dissipation, uncommon for the other flow configurations. Table~\ref{tbl:simulation_parameters} lists the representative cases used for the energy-profile figure. The simulations can generally be classified as simple shear, with $G=0$ and $U\ne 0$, shear with gravity with $G\ne0$ and $U\ne0$, gravity-driven flow with $G\ne0$ and $U=0$, and diverging shear with $G\ne0$ and $U\ne0$.

\begin{table}
\caption{Representative simulations input parameters for the cases shown in Fig.~\ref{fig:more_energy_profiles}.}
\label{tbl:simulation_parameters}
\centering
\setlength{\tabcolsep}{4pt}
 \begin{tabular}{|c | l | c | c | c | c | c |} 
 \hline
 Case & Flow & $P$ & $U$ & $G$ & $\psi$ & $N$ \\ [1ex]
 \hline
 1 & simple shear & $2\times10^{-6}$ & $7.37646\times10^{-7}$ & $0$ & -- & $20{,}754$ \\
 2 & simple shear & $5\times10^{-6}$ & $2.6637\times10^{-5}$ & $0$ & -- & $20{,}754$ \\
 3 & simple shear & $1\times10^{-5}$ & $1.530808\times10^{-2}$ & $0$ & -- & $20{,}754$ \\
 4 & simple shear & $3\times10^{-6}$ & $4.4\times10^{-2}$ & $0$ & -- & $20{,}754$ \\
 5 & shear with gravity & $2\times10^{-6}$ & $6\times10^{-5}$ & $2.5\times10^{-9}$ & $90^{\circ}$ & $42{,}134$ \\
 6 & shear with gravity & $2\times10^{-6}$ & $4\times10^{-5}$ & $2\times10^{-8}$ & $90^{\circ}$ & $42{,}134$ \\
 7 & gravity driven & $2\times10^{-6}$ & $0$ & $3\times10^{-8}$ & $90^{\circ}$ & $42{,}134$ \\
 8 & gravity driven & $5\times10^{-6}$ & $0$ & $8\times10^{-8}$ & $60^{\circ}$ & $42{,}134$ \\
 9 & diverging shear & $5\times10^{-6}$ & $2.6637\times10^{-5}$ & $1.24\times10^{-7}$ & $0^{\circ}$ & $20{,}754$ \\
 10 & shear with gravity & $2\times10^{-6}$ & $1\times10^{-4}$ & $2.5\times10^{-8}$ & $90^{\circ}$ & $42{,}134$ \\
 [1ex] 
 \hline
 \end{tabular}
\end{table}

All simulations are brought to a statistically stationary state by inspecting that the volume-averaged energy production ratio $\dot E\equiv \int_\text{height}(\sigma_{xy}\dot\gamma - \chi )dy/\int_\text{height}(\sigma_{xy}\dot\gamma + \chi )dy$ is less than 10\% when averaged over 100 time steps. Then, data is collected at an interval ranging from 20,000 to 100,000 time steps, for at least 1,000 frames. We note that $\dot E$ only indicates the steadiness of the entire volume, so it can be reasonable, even if $\dot E\ne0$, for subvolumes of the flowing material to be approximately steady. In practice it is not possible to rule out very slow changes to the macroscopic flow state resulting from, for example, size-based segregation or diffusion from a distant unsteady region. Fig.~\ref{fig:calibration}c shows an example time series of $\dot E$.

\begin{figure}
\includegraphics[width=0.99\textwidth]{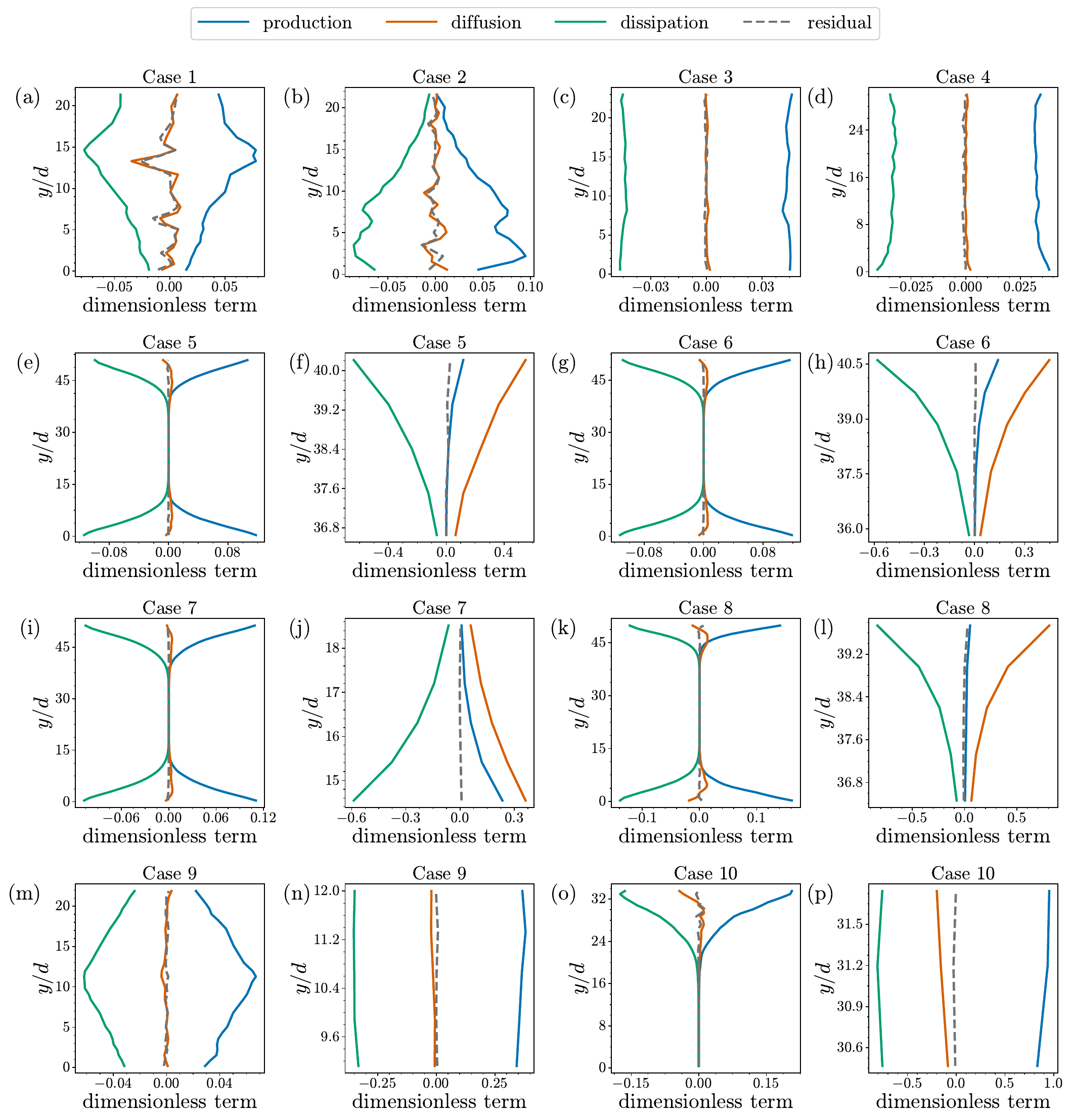}
\caption{\label{fig:more_energy_profiles}Examples of spatial profiles of each term in the granular temperature evolution equation. Each term is nondimensionalized by the $y$-average of $\chi$. }
\end{figure}

In the main text we plot spatial profiles of each term in the granular temperature evolution equation. Here in Fig.~\ref{fig:more_energy_profiles} we include more examples. Some panels are zoom-ins of others, demonstrating that certain regions may be dominated by diffusion even when the bulk is dominated by shear, and diffusion can add or subtract $T$ depending on the local balance. We note that the sign of $\frac{dQ_y}{dy}$ indicates whether a material point is experiencing influx or efflux of granular temperature. Data with $\frac{dQ_y}{dy}<0$ is less prominent for the flows under consideration, associated with small regions typically near the boundaries. With the assumption that the local values of $\mu_0$ and $\Theta_0$ are functions of $I$ alone (main text), we can think of material points as ``cooled" or ``warmed" compared to the local state for a given value of $I$, with the ``warmed" regions, at low $I$, constituting the regions that flow with $\mu<\mu_s$ because of the input of fluidizing fluctuations (that they receive from ``cooled" regions). The observation~\cite{kim2020power} that, for constant $I$, $\mu$ and $\Theta$ trend oppositely, means $\alpha>(<)0$ suggests $\mu<(>)\mu_0$ and $\Theta>(<)\Theta_0$, a trend that is satisfied in part by our data. Deviations from this trend could imply that $\mu_0$ and $\Theta_0$ are not functions of $I$ alone, that the fitted functions $\mu_0(I)$ and $\Theta_0(I)$ lose some accuracy, or that the gradient of $\Theta$ alone is not a sufficient measure of nonlocality. A comparison of the predictive power of our model to others (Sec.~\ref{sec:model_predictions}) suggests that these potential limitations are not especially consequential for the flows considered herein.

For some cases like Fig.~\ref{fig:more_energy_profiles}(a,b), the ensemble-averaged data has grain-scale spatial fluctuation, despite satisfying our criterion for steadiness of the granular temperature, possibly owing to flow instabilities. We also note that the homogeneous shear tests in general are not perfectly homogeneous, although their deviations from homogeneity are small compared to those of the intentionally heterogeneous flows and we observe they have negligible diffusive flux of $T$. In reality, wall effects induce some heterogeneity, in a way that can be described~\cite{miller2013eddy} by analogy to turbulence. 

\subsection{Coarse-graining procedure}
\label{sub:coarse_graining_procedure}
To coarse-grain prior to ensemble-averaging, we employ two approaches. The primary approach is grain volume averaging~\cite{zhang2017microscopic,kim2020power}, applied to individual grains as control volumes. To ensemble-average the grain-averaged quantities, we compute streamline averages for instantaneous values for streamlines every $0.1d$ along $y$, which are then averaged temporally (over well-separated snapshots) for statistically stationary flows. The resulting profiles are still typically noisy and benefit from further spatial smoothing over a neighborhood no greater than several $d$~\cite{zhang2017microscopic}. We find in some flow regions like boundary layers or other regions of high gradients that a neighborhood greater than $d$ smooths over important features, whereas in some cases a larger neighborhood of $3d$ is helpful for the large noise. To reconcile this, we utilize a Haar adaptive binning. Beginning with a grid of bin size $3d$, we estimate the noise in a bin for the velocity (though apply the same resulting grid to all fields) as the median absolute deviation. We divide a bin into two if the ratio of absolute difference of the two sides' means to their noise is greater than $\sqrt{2\log N_\text{points}}$, keeping though a minimum bin size of $0.5d$. In any case, larger simulation domain volumes would be helpful in reducing the need for clever approaches to the ensemble average. The result of all of this temporal and spatial averaging is (approximately) the ensemble-average operator $\langle .\rangle$, defined on a streamline of location $y$ and size $S$ for a field $f$ with per-grain values $f_g$ and overlap sizes $S_g$ as
\begin{equation}
\label{eqn:cgoperator}
\langle f\rangle(y) = \text{time and neighborhood average of} \left[\frac1S\sum_\text{overlapping grains} f_gS_g\right].
\end{equation}
Sizes $S$ and $S_g$ are areas. The total streamline size $S$ is $L_xL_z$ for extensive quantities (velocities, temperature, etc.) and the sum $\sum_\text{overlapping grains}S_g$ for intensive quantities (stresses, energy fluxes, dissipation rate, and packing fraction).

Given contact force vector $\boldsymbol F$, location of contact $\boldsymbol r$ relative to grain center, velocity vector $\boldsymbol u_g$, and grain radius $r_g$, we use the following microscopic (per-grain) definitions for fluctuating velocity vector $\boldsymbol u'_g$, stress tensor $\boldsymbol\sigma_g$ (with contributions from fluctuating velocities $\boldsymbol\sigma_g^\text{kin}$ and from contact forces $\boldsymbol\sigma_g^\text{con}$), fluctuating energy flux vector $\boldsymbol Q_g$, and kinematic dissipation rate $\chi_g$ (for contacting grain $c$):
\begin{subequations}
\begin{equation}
\boldsymbol u'_g = \boldsymbol u_g - \langle \boldsymbol u_g \rangle,
\end{equation}
\begin{equation}
\boldsymbol\sigma_g = -\rho \boldsymbol u'_g\otimes\boldsymbol u'_g - \frac1{\frac43\pi d^3}\sum_\text{contacts}(\boldsymbol F\otimes \boldsymbol r)_\text{contact} \equiv \boldsymbol\sigma_g^\text{kin} + \boldsymbol\sigma_g^\text{con},
\end{equation}
\begin{equation}
\boldsymbol Q_g = \left(\boldsymbol\sigma_g^\text{con}-\frac12\rho\boldsymbol u'_g\otimes\boldsymbol u'_g\right)\boldsymbol\cdot\boldsymbol u'_g,
\end{equation} 
\begin{equation}
\chi_g = \frac1{\frac83\pi d^3}\sum_\text{contacts}\boldsymbol F\boldsymbol\cdot(\boldsymbol v_c - \boldsymbol v_g).
\end{equation}
\end{subequations}
Note that this form of the dissipation rate~\cite{pahtz2015fluctuation} can be positive or negative, and is not generally equal to $\sum_cW_c$. At the per-grain-scale in our formulation, packing fraction $\phi_g=1$. These per-grain values are then fed to Eq.~\ref{eqn:cgoperator} to compute coarse-grained fields. We note that ensemble-averaging implies the removal of all fluctuations, although some have argued that the spatial average alone (not the steady-state temporal average) should be used when defining a fluctuating quantity~\cite{zhang2017microscopic}. The conclusions drawn from our data are unaffected by this choice. It is possible that long-lasting unsteadiness could cause a drift in the temporal average that requires a more complicated ensemble-averaging procedure, such as one including temporal filtering. 

\begin{figure}
\includegraphics[width=0.99\textwidth]{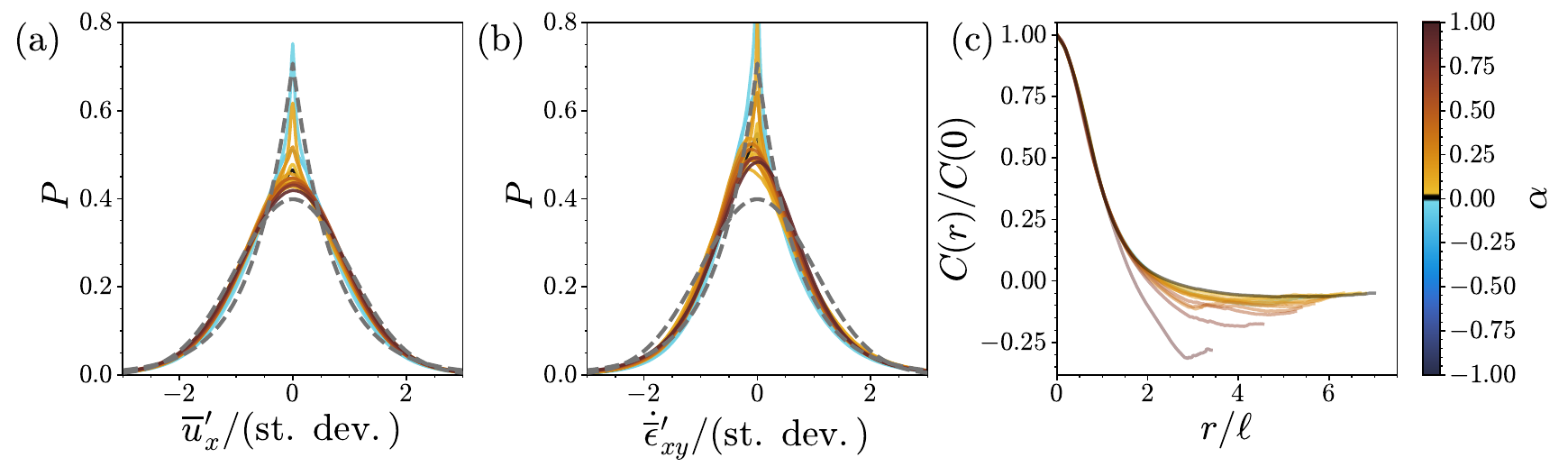}
\caption{\label{fig:distributions}Probabilistic distributions of the coarse-grained $x$ component of the velocity fluctuation (a) and the $xy$ component of the strain rate fluctuation (b) scaled by their standard deviations, across all steady simulation streamlines, averaged into bins of varying relative diffusion $\alpha$. Gaussian and Laplacian distributions are plotted for comparison. (c) The correlation function divided by the $e$-fold distance that defines the correlation length $\ell$.}
\end{figure}

The grain-average method of coarse-graining cannot be used to measure local strain-rates. Therefore, we coarse-grain a subset of our data with a filtering approach (e.g.~\cite{zhang2010coarse,goldhirsch2010stress,weinhart2013coarse}), and note the ensemble-average is then the simple average along homogeneous directions and in time of the continuum quantity (computed on a uniform grid for simplicity). Defining, for relative position $\boldsymbol r_\text{rel}$, a filter $G(\boldsymbol r_\text{rel}; \Delta)=(\sqrt{2\pi}\Delta)^{-3}\exp(-||(\boldsymbol r/\Delta)||^2/2)$ and summing across grains $g$, we have continuum fields of packing fraction
\begin{equation}
\overline{\phi} = \frac1\rho \sum_gm_gG(\boldsymbol r - \boldsymbol r_g),
\end{equation}
velocity
\begin{equation}
\overline{\boldsymbol u}(\boldsymbol r) = \frac{ \sum_g m_g\boldsymbol u_gG(\boldsymbol r - \boldsymbol r_g)}{\rho\overline{\phi}},
\end{equation}
noting $\boldsymbol u_g'=\boldsymbol u_g-\overline{\boldsymbol u_g}$ and summing also across contacts $c$, stress
\begin{equation}
\label{eqn:cgstress}
\overline{\boldsymbol \sigma} = -\frac12\sum_{g,c}\boldsymbol F_c\otimes \boldsymbol \ell_c \int_0^1ds G(\boldsymbol r-\boldsymbol r_g + s\boldsymbol\ell_c) - \sum_gm_g\boldsymbol u_g'\otimes\boldsymbol u_g'G(\boldsymbol r- \boldsymbol r_g)\equiv \overline{\boldsymbol\sigma^\text{con}} + \overline{\boldsymbol\sigma^\text{kin}},
\end{equation}
and velocity gradient
\begin{equation}
\overline{\nabla\boldsymbol u} = \sum_g\frac{m_g\boldsymbol u_g\otimes\left(\nabla G(\boldsymbol r-\boldsymbol r_g\right)\overline{\phi} + G(\boldsymbol r-\boldsymbol r_g)\overline{\nabla\phi})}{\rho\overline{\phi}^2}.
\end{equation}
The overline $\overline f$ and the subscript $f_g$ indicate coarse-grained quantities, for the two different approaches considered here. The fluctuations of velocity in Eq.~\ref{eqn:cgstress} are from the coarse-grained field rather than the ensemble-average, so the entire expression plays the role of the grain-scale contact stress for ensemble-averaging the equations. We compute these fields, choosing $\Delta=d/2$ (note for either method we choose the length scale to generate coarse-grain fields corresponding to the grain-scale), to measure the distribution functions of $\overline{u}_x$ and $\dot\epsilon_{xy}=\frac12(\partial \overline{u}_y/\partial x + \partial \overline{u}_x/\partial y)$ and the correlation function $C(r,y)\propto\langle\overline{\boldsymbol u}(\boldsymbol x,t)
\boldsymbol\cdot\overline{\boldsymbol u}(\boldsymbol x+\boldsymbol r,t)\rangle$ in Fig.~\ref{fig:distributions}. The distributions and correlation functions are both computed per streamline (using $\boldsymbol r\perp\boldsymbol e_y$) and then binned by $\alpha$, and the correlation length $\ell(y)$ is defined as the (interpolated) distance for the normalized $C(\boldsymbol r,y)$ to fall by a factor of $e$.

The choice of coarse-graining method is consequential when considering mesoscopic scales. Figure~\ref{fig:more_energy_profiles} demonstrates that our grain-averaging method is sufficiently accurate for our study. One could take the limit $\Delta\to0$, turning the Gaussian to a Dirac distribution kernel, as done in Ref.~\cite{pahtz2015fluctuation}, which then requires defining control volumes to alleviate the singularities. One could also take coarse-grained stress tensor $\overline{\boldsymbol\sigma}$, and via Favre-averaging identify the dissipation rate $\chi=\langle\overline{\boldsymbol\sigma}:\overline{\nabla\boldsymbol u}\rangle$. However, one must be careful that the balances (i.e. Fig.~\ref{fig:more_energy_profiles}) are not sensitive then to the specific coarse-graining kernel. Furthermore, Ref.~\cite{pahtz2015fluctuation}, with their Dirac distribution method, found $\chi$ to be too noisy and instead inferred $\chi$ from the equation residual. We do not encounter this problem for the grain-averaging method. In any case the appropriateness of point-particle coarse-graining methods~\cite{goldhirsch2010stress} for characterizing the mesoscale requires further investigation.

\section{Scaling relations}
We used \verb|scipy.optimize.curve_fit| and \verb|numpy.polyfit| for fitting curves to our data, with all parameters and their error estimates shown in Table~\ref{tbl:scaling_fit_parameters}. The local friction rheology 
\begin{equation}
\mu_0(I) = \mu_s + \frac{\mu_2-\mu_s}{\frac{I^\star}I + 1}
\label{eqn:mu0}
\end{equation}
and the local dimensionless temperature equation
\begin{equation}
\Theta_0(I) = A_\Theta I^{n_\Theta}
\label{eqn:Theta0}
\end{equation}
were fit using only data from homogeneous shear tests. The conductivity relation 
\begin{equation}
\tilde\kappa(\Theta) = A_\kappa\Theta^{-a_\kappa}
\label{eqn:kappaTheta}
\end{equation}
was fit using only the nonhomogeneous data (with $|\alpha|>0.01$). The friction-density relation 
\begin{equation}
\mu(\phi) = \mu_\phi - \frac{A_\mu}{\phi_\mu-\phi}
\end{equation}
was fit using all steady data. The power law exponents $a_\eta$, $a_\mu$, $a_\chi$, and $a_\kappa$ were estimated by dividing the data into a uniform 40-bin grid over the range $10^{-7}\le I\le10^{-2}$ and fitting a power law $f\approx\Theta^{-a_f}$ (as done by Ref.~\cite{kim2020power} for $\mu\propto\Theta^{-a_\mu}$) over all bins with at least six points. Their 10th and 90th percentiles across the retained bins are reported as the error estimate. We note there could be further, physically meaningful dependence $a_f(I)$ if one desires greater accuracy. This procedure identified $f_\kappa\approx\text{const.}$ (main text Fig.~3(g)), motivating our simple model $\tilde\kappa\approx A_\kappa\Theta^{-a_\kappa}$ for which we measure the best-fitting $A_\kappa$ from the data after determining $a_\kappa$.

\begin{figure}
\includegraphics[width=0.99\textwidth]{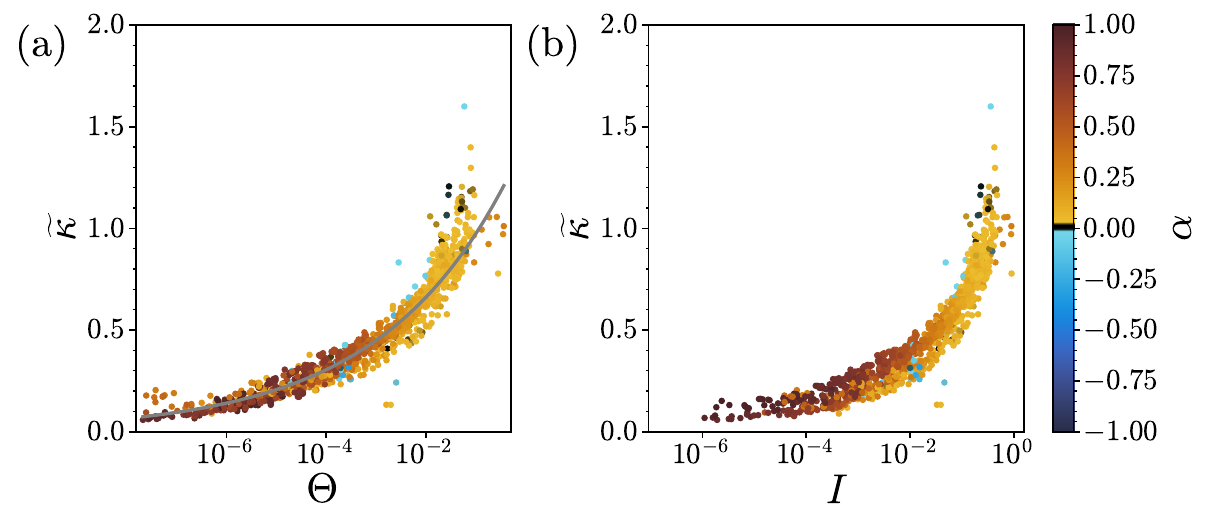}
\caption{\label{fig:kappa_theta_I} Conductivity models, colored by $\alpha$. (a) Dimensionless conductivity $\tilde\kappa$ versus dimensionless granular temperature $\Theta$, with the fit $\tilde\kappa(\Theta)=A_\kappa\Theta^{-a_\kappa}$, as used in this work. (b) A reasonable alternative could relate $\tilde\kappa=\tilde\kappa(I)$.}
\end{figure}

Note for the relations $\mu\Theta^{a_\mu}=f_\mu(I)$, $\tilde\eta\Theta^{a_\eta}=f_\eta(I)$, since $f_\mu(I)=\mu_0\Theta_0^{a_\mu}$ and $\mu=\eta\dot\gamma/p=\tilde\eta I/\sqrt\Theta$, we find $a_\eta=a_\mu-\frac12$ and $f_\eta(I)=\mu_0\Theta_0^{a_\mu}/I$. (We measure $a_\mu$ and $a_\eta$ both from fitting $\mu$ and $\tilde\eta$ as described above, then average the two measurements.) Furthermore, for $\tilde\chi\Theta^{a_\chi}=f_\chi(I)$, since the homogeneous balance gives $\mu_0 I = \tilde\chi(I,\Theta_0)\sqrt{\Theta_0}$, we find $f_\chi(I)=\mu_0\Theta_0^{a_\chi-\frac12}I$, meaning the model is calibrated once $\mu_0(I)$, $\Theta_0(I)$ (both of which are determined by simple shear tests), $A_\kappa$, $a_\kappa$, $\mu(\phi)$, $a_\eta$, and $a_\chi$ are known. 

We also measured the distinct cooperativity lengths
\begin{equation}
\xi_\text{NGF}^2=\frac{(\mu_s-\mu)g+\dfrac{\mu_2-\mu}{I^\star}\sqrt{\dfrac{\rho d^2}{p}}\,\mu g^2}{|\mu-\mu_s|d^2g/dy^2},
\end{equation}
which we take from the equation~\cite{kamrin2014nonlocal}
\begin{equation}
A_\text{NGF}^2d^2\frac{d^2g}{dy^2} - (\mu_2-\mu_s)\frac{\mu_s-\mu}{\mu_2-\mu}g - \frac{\mu_2-\mu_s}{I^\star}\sqrt{\frac{\rho d^2}p}\mu g^2 = 0,
\label{eqn:NGF_full}
\end{equation}
and
\begin{equation}
\xi_\text{EKT}^2=\frac{g-\dot\gamma^2p/\chi}{d^2g/dy^2},
\end{equation}
omitting points with $|\alpha|<0.01$ or negative values in the square roots. The latter, previously predicted to be constant~\cite{berzi2024granular}, is not constant for our data (see Tbl.~\ref{tbl:scaling_fit_parameters} and main text). Using the typical~\cite{kamrin2014nonlocal} fitting function
\begin{equation}
\xi_\text{NGF}=A_\text{NGF}d\sqrt{\frac{\mu_2-\mu}{(\mu_2-\mu_s)|\mu_s-\mu|}},
\end{equation}
we measured $A_\text{NGF}$ as the simulation-suite mean reported in Tbl.~\ref{tbl:scaling_fit_parameters}. Its spread (main text) and error estimate (Tbl.~\ref{tbl:scaling_fit_parameters}) suggest it is also not a constant as previously predicted~\cite{kamrin2014nonlocal}. 

\begin{figure}
\includegraphics[width=0.65\textwidth]{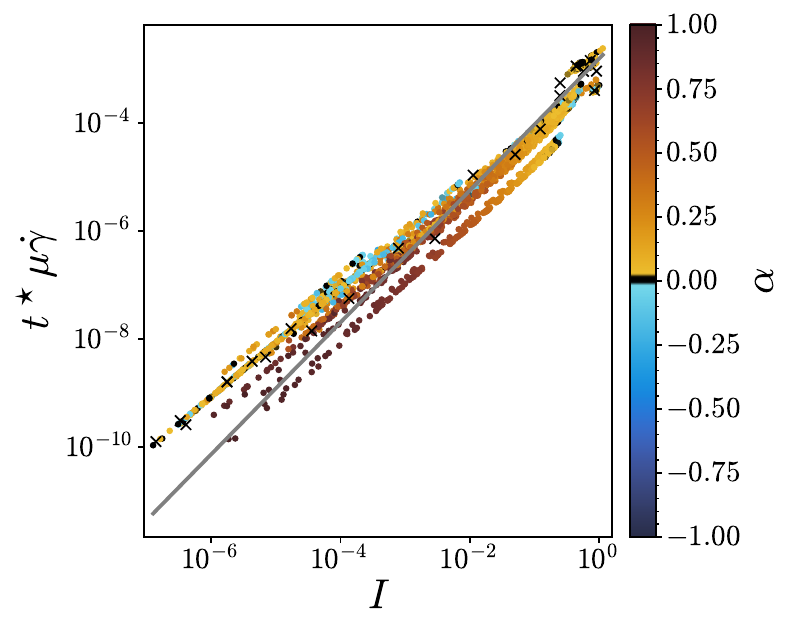}
\caption{\label{fig:production_Irmer_fit.pdf}Best-fitting line for the production model proposed by Ref.~\cite{irmer2025granular}.}
\end{figure}

Ref.~\cite{irmer2025granular} proposed a steady heat equation (HE) for $\Theta$
\begin{equation}
0=D_\text{HE}d^2\frac{d^2\Theta}{dy^2}-B_\text{HE}\Theta+A_\text{HE}I^{a_\text{HE}}
\label{eqn:heat_equation}
\end{equation}
for which we compute the parameters (requiring the assumption of uniform $p$)
\begin{equation}
t^\star\equiv d\sqrt{\frac{\rho}{E}},\qquad
D_\text{HE}=\frac{-t^\star dQ_y/dy}{p d^2\left(d^2\Theta/dy^2\right)},\qquad
A_\text{HE}I^{a_\text{HE}}=t^\star\mu\dot\gamma,\qquad
B_\text{HE}=\frac{t^\star\chi}{p\Theta}
\end{equation}
presented in Tbl.~\ref{tbl:scaling_fit_parameters} as simulation-sweep averages, grouping by bins in $\Theta$ before averaging to prevent density of data points from skewing the average, except for $A_\text{HE}$ and $a_\text{HE}$ which are determined by curve-fitting the production term (Fig.~\ref{fig:production_Irmer_fit.pdf}). These too show considerable spread. We note the parameter $A_\text{HE}$, assumed constant, presents a physical inconsistency because the rheological model $\mu(I,\Theta)$ is used for the momentum balance but not the production $\mu p \dot\gamma\to A_\text{HE}I^{a_\text{HE}}$. 

\begin{table}
\caption{Parameter fits for our model and others~\cite{kamrin2014nonlocal,berzi2024granular,irmer2025granular}. Error estimates are either $\pm$ one standard deviation or $(10^\text{th},90^\text{th})$ percentiles.}
\label{tbl:scaling_fit_parameters}
\centering
\setlength{\tabcolsep}{4pt}
\renewcommand{\arraystretch}{0.84}
\begin{tabular}{|l | c | c | l |}
\hline
Relation & Parameter & Fit value & \makebox[0.34\textwidth][l]{Error estimate} \\ [1ex]
\hline
Exponents & $a_\eta$ & $-0.3973$ & $(-0.4929, -0.3289)$ \\
 & $a_\mu$ & $0.1027$ & $(0.0071, 0.1711)$ \\
 & $a_\chi$ & $0.1701$ & $(-0.01598, 0.4183)$ \\
 & $a_\kappa$ & $-0.1679$ & $(-0.2791, -0.03436)$ \\
\hline
$\mu_0(I)$ & $\mu_s$ & $0.2999$ & $\pm 0.0016$ \\
 & $\mu_2$ & $0.720$ & $\pm 0.021$ \\
 & $I^\star$ & $0.596$ & $\pm 0.061$ \\
\hline
$\Theta_0(I)$ & $A_\Theta$ & $0.488$ & $\pm 0.071$ \\
 & $n_\Theta$ & $1.840$ & $\pm 0.016$ \\
\hline
$\widetilde{\kappa}(\Theta)$ & $A_\kappa$ & $1.4319$ & $\pm 0.0061$ \\
\hline
$\mu(\phi)$ & $\mu_\phi$ & $0.808$ & $\pm 0.028$ \\
 & $A_\mu$ & $0.0687$ & $\pm 0.0093$ \\
 & $\phi_\mu$ & $0.747$ & $\pm 0.010$ \\
\hline
Other models & $A_{\mathrm{NGF}}$ (Ref.~\cite{kamrin2014nonlocal}) & $0.5522$ & $(0.3156, 0.8337)$ \\
 & $\frac{\xi_{\mathrm{EKT}}}{d}$ (Ref.~\cite{berzi2024granular}) & $1.903$ & $(1.104, 2.293)$ \\
 & $D_{\mathrm{HE}}$ (Ref.~\cite{irmer2025granular}) & $0.0331$ & $(0.006084, 0.2293)$ \\
 & $A_{\mathrm{HE}}$ (Ref.~\cite{irmer2025granular}) & $1.549\times10^{-3}$ & $\pm 0.023\times10^{-3}$ \\
 & $a_{\mathrm{HE}}$ (Ref.~\cite{irmer2025granular}) & $1.220$ & $\pm 0.026$ \\
 & $B_{\mathrm{HE}}$ (Ref.~\cite{irmer2025granular}) & $0.6775$ & $(0.005893, 191.6)$ \\
\hline
\end{tabular}
\end{table}

\begin{figure}
\includegraphics[width=0.65\textwidth]{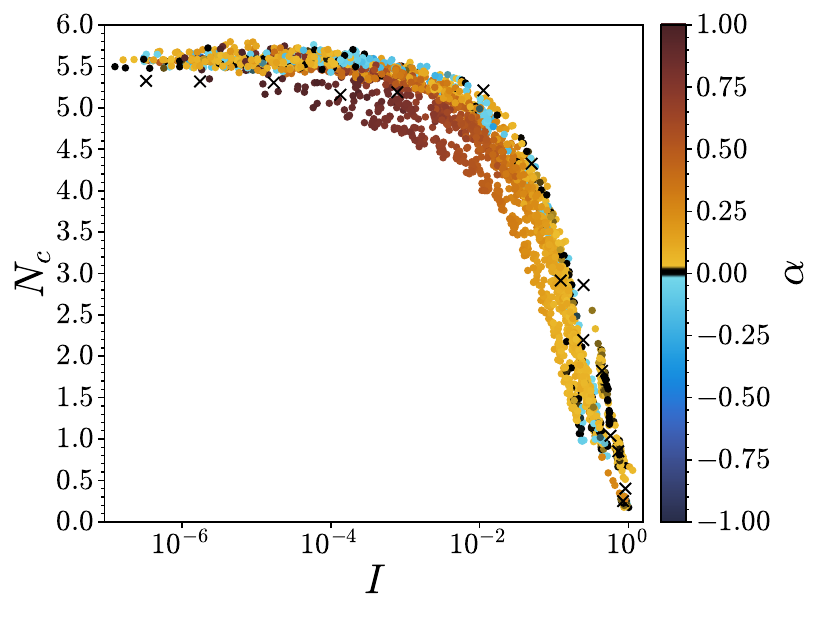}
\caption{\label{fig:Nc_vs_I}Scatterplot of coordination number $N_c$ versus inertial number $I$ for our simulations.}
\end{figure}

\begin{figure}
\includegraphics[width=0.99\textwidth]{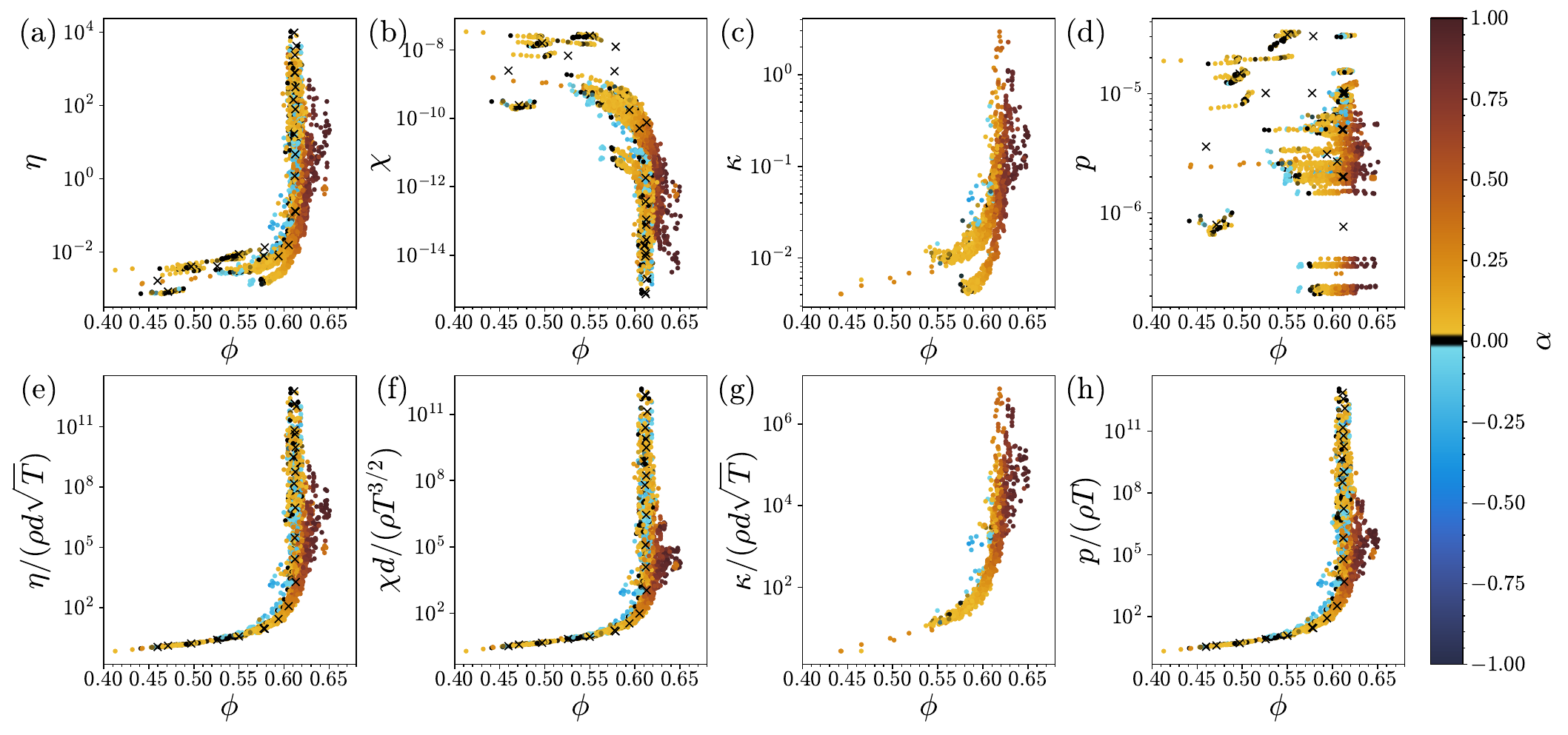}
\caption{\label{fig:unscaled}Dimensional (a-d) and alternatively scaled dimensionless (e-h) viscosity $\eta$ (a,e), dissipation rate $\chi$ (b,f), conductivity $\kappa$ (c,g), and pressure $p$ (d,h). The alternate scaling compared to that used in our model replaces $p$ with $\rho T$~\cite{brilliantov2010kinetic}.}
\end{figure}

Regarding the applicability of the known kinetic theory scalings, we scatter the coordination number $N_c$ vs. $I$ in Fig.~\ref{fig:Nc_vs_I}, showing that the binary collision assumption is generally not valid for our system. The typical stress scale in kinetic theory, $\rho T$, is then not sufficient to collapse our data. The uncollapsed dimensional transport coefficients $\eta$, $\kappa$, and $\chi$, as well as the pressure $p$, are shown vs. $\phi$ in Fig.~\ref{fig:unscaled}(a-d), and their alternative dimensionless values, scaled by $\rho T$ and $d$, are shown in Fig.~\ref{fig:unscaled}(e-h). The dimensional quantities are spread over decades, and their alternative dimensionless values only appreciably collapse in the low-$\phi$ regime. Instead, using $p$ as a stress scale (main text) is a considerable improvement, handling both low $\phi$ (where $p\propto\rho T$) and high $\phi$, where there is no obvious collapse for $p(\phi,\rho T)$. We also note that the kinetic theory model of Ref.~\cite{berzi2015different} describes flows with enduring contacts but some of its theoretical predictions such as $\tilde\eta=\text{const.}$ are in qualitative disagreement with our data. 

As discussed in the main text, the $\phi$-based constitutive models still lose accuracy in the dense regime and are improved by the power-law based scalings. One possible reason for this discrepancy is the nonuniversality of the distributions of fluctuating motions, plotted in Fig.~\ref{fig:distributions} which clearly depend in form upon relative diffusion $\alpha$. Our collapses are phenomenological evidence that this feature is reflected at least in part by the ($I$,$\Theta$)-based power law scalings, and a theoretical explanation for this is the subject of future investigation. 

\section{Details of the derivation of the fluidity equation}
As explained in the main text, for the NGF definition $g_0=I_0\sqrt{p/\rho d^2}/\mu$ we linearize first for $I\to0$. Note, using our constitutive assumptions, $\mu I - \tilde\chi\sqrt\Theta = \mu^2 g \sqrt{\frac{\rho d^2}p}[1-(\mu/\mu_0)^{a^\star}]$ with $a^\star=(a_\chi-a_\mu-\frac12)/a_\mu$. In the limit $I\to0$, noting $g_0=0$, we identify $\mu I - \tilde\chi\sqrt\Theta = \sqrt{\frac{\rho d^2}p}[(\mu/\mu_s)^{a^\star}-1](g_0-g)$. For $I=I_0+\delta I,~\delta I\ll I_0$, linearization and application of the definition $\mu_0(I_0)=\mu$ gives $\mu I - \tilde\chi\sqrt\Theta=-\mu^2g\sqrt{\frac{\rho d^2}p}a^\star\mu^\prime(I_0)(g_0-g)$. In both limits, our simulations confirm the prefactor to $g_0-g$ is positive (note $a^\star$ is negative).

For the diffusion term, we neglect $\frac{d\mu}{dy}$, $\frac{d\mu_0}{dy}$, and $\frac{dp}{dy}$ to get 
\begin{equation}
\frac1{\sqrt\Theta}\tilde\kappa\frac{d\Theta}{dy}\approx \sqrt{\frac{\rho d^2}{p}}\mu\sqrt{\left(\frac{\mu_0}{\mu}\right)^{a_\mu}\Theta_0}(\log\Theta)^\prime\tilde\kappa\frac{dg}{dy}.
\end{equation}
Further neglecting the gradient of the prefactor to $\frac{dg}{dy}$ allows us to identify 
\begin{equation}
\frac{\xi^2}{d^2}=\frac1{C}\sqrt{\frac{\rho d^2}p}\mu\sqrt{\left(\frac{\mu_0}\mu\right)^{a_\mu}\Theta_0}(\log\Theta_0)^\prime\tilde\kappa,
\end{equation}
where $C$ is the prefactor to the $(g_0-g)$ terms derived above. Note from Eq.~\ref{eqn:Theta0} that $(\log\Theta_0)'=\frac{n_\Theta}{I}$.

\section{Model predictions of flow fields}
\label{sec:model_predictions}
For extensive comparison of the predictions of some models of dense granular flows, see Ref.~\cite{debnath2022comparison}. We here summarize and demonstrate how one might use our model to solve for a flow field, with comparison to DEM and the models of Ref.~\cite{kamrin2014nonlocal} and~\cite{irmer2025granular}. The generic transport equations for ensemble-averaged mass, momentum, and granular temperature, dropping the $\langle.\rangle$ notation for brevity (and referring to constant grain density as $\rho$), and defining convective derivative $D_t=\frac{\partial }{\partial t}+\boldsymbol u\boldsymbol\cdot\nabla$, are
\begin{subequations}
\begin{equation}
    D_t\phi = -\phi\nabla\boldsymbol\cdot\boldsymbol u,
\end{equation}
\begin{equation}
    \rho\phi D_t\boldsymbol u = \nabla\boldsymbol\cdot\boldsymbol\sigma + \rho \phi \boldsymbol G,
\end{equation}
\begin{equation}
    \rho\phi D_t T = \boldsymbol\sigma:\nabla\boldsymbol u-\nabla\boldsymbol\cdot\boldsymbol Q - \chi.
\end{equation}
\end{subequations}
These equations describe flows that can be unsteady and multidirectional, and may even require solving an angular momentum balance (not shown) for the stress asymmetry~\cite{goldhirsch2010stress}. Unsteadiness is known to affect constitutive models. Calibrating constitutive models for unsteady settings is not trivial and requires further investigation, but it is common~\cite{yuan2026toward} to simply apply rheologies calibrated from steady data to unsteady flows. Note that unsteadiness is possible even if $\partial f/\partial t=0$, because of the convective term which can transport material from one neighborhood to another in finite time. The steady unidirectional simplification trivializes the mass balance, and the momentum and granular temperature equations become
\begin{subequations}
\begin{equation}
    \frac{d\sigma_{yy}}{dy} + \rho\phi G_y = 0,~~~~\frac{d\sigma_{zz}}{dz}=0,~~~~\frac{d\sigma_{xy}}{dy} + \rho\phi G_x = 0,
\end{equation}
\begin{equation}
    \sigma_{xy}\dot\gamma - \frac{dQ_y}{dy} - \chi = 0.
\end{equation}
\end{subequations}
For unidirectional granular flows it is common to approximate $\sigma_{yy}\approx p=\frac13(\sigma_{xx}+\sigma_{yy}+\sigma_{zz})$, or to calibrate (as we have done) rheological models using $\sigma_{yy}$ as the stress scale, eliminating either way the need to determine $\sigma_{zz}$ or $\sigma_{xx}$. Without this approximation, further constitutive assumptions would be required to determine each of the normal stresses. We proceed then, including a condition for $\phi$:
\begin{subequations}
\begin{equation}
    \frac{d}{dy}(\eta\dot\gamma) + \rho\phi G_x = 0,
\end{equation}
\begin{equation}
    \eta\dot\gamma^2 + \frac{d}{dy}\left(\kappa\frac{dT}{dy}\right) - \chi = 0,
    \label{eqn:generic_temperature_equation}
\end{equation}
\begin{equation}
    p(y) = \int\rho\phi G_y,~~~~\frac{\eta\dot\gamma}{p} = \mu(\phi).
\end{equation}
\end{subequations}
We note that assuming a linear profile of $p$ under gravity as sometimes done~\cite{kim2020power} would imply uniform $\phi$, and therefore uniform $\mu$ by our model $\mu(\phi)$. Models that lack a constitutive model for $\phi$ then can only make predictions for flows of uniform $\phi$ unless the stress field is already known (see Ref.~\cite{debnath2022comparison} for further discussion). Our system of ODEs requires boundary conditions. We take conditions on $u_x$, $T$, and $\int_{y_b}^{y_t}\phi dy$ from DEM because our coarse-graining procedure does not include the rigid walls~\cite{weinhart2013coarse}, which would be an added complication requiring constitutive modeling of the boundary interactions (an active area of research~\cite{debnath2022different}). The boundaries here correspond to bottom/top material points at $y_b$ and $y_t$ near the rigid walls.

\begin{figure}
\includegraphics[width=0.99\textwidth]{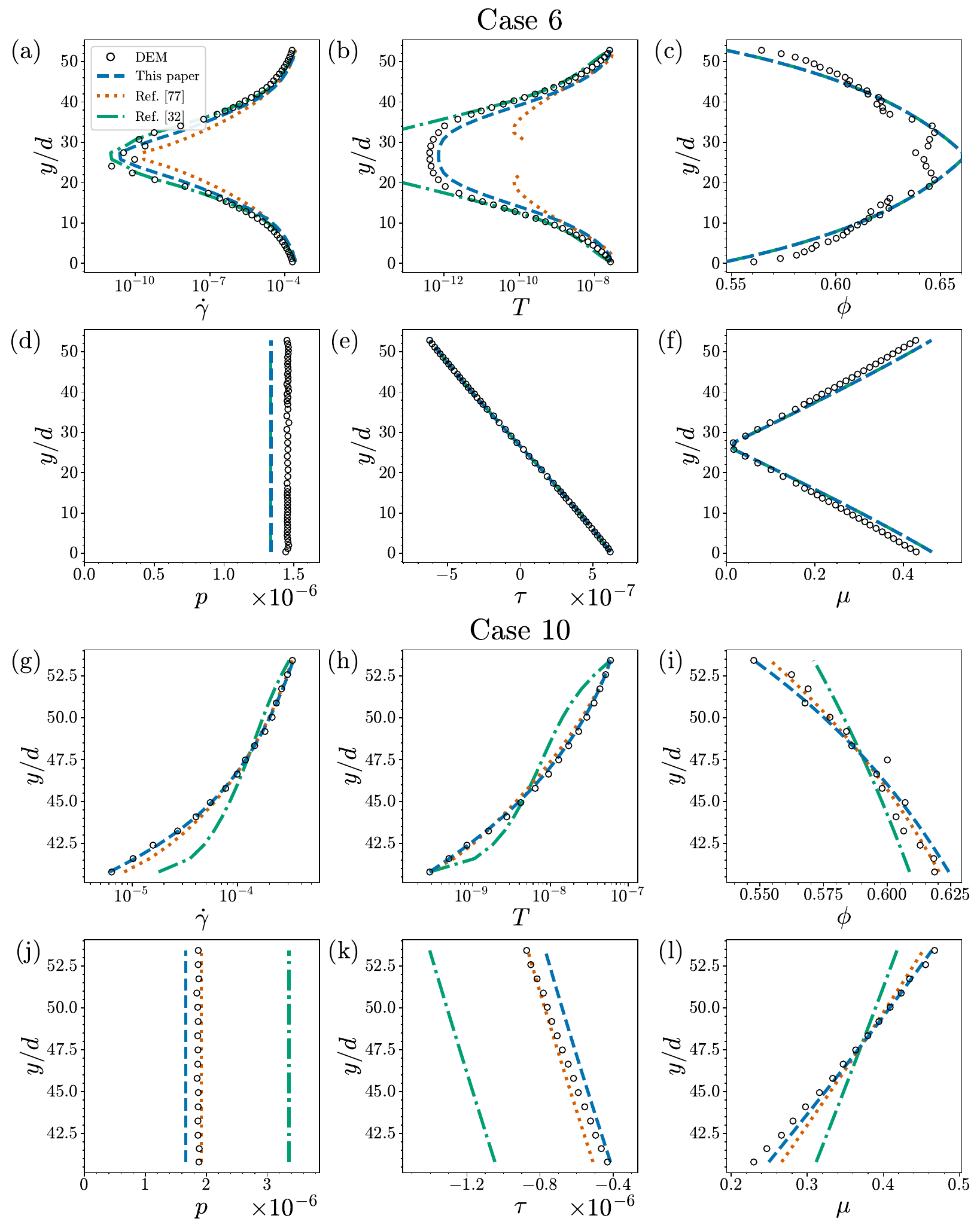}
\caption{\label{fig:model_solutions}Example solutions from our model for two cases from Tbl.~\ref{tbl:simulation_parameters}, compared to Refs.~\cite{kamrin2014nonlocal,irmer2025granular} and our DEM.}
\end{figure}

\begin{figure}
\includegraphics[width=0.99\textwidth]{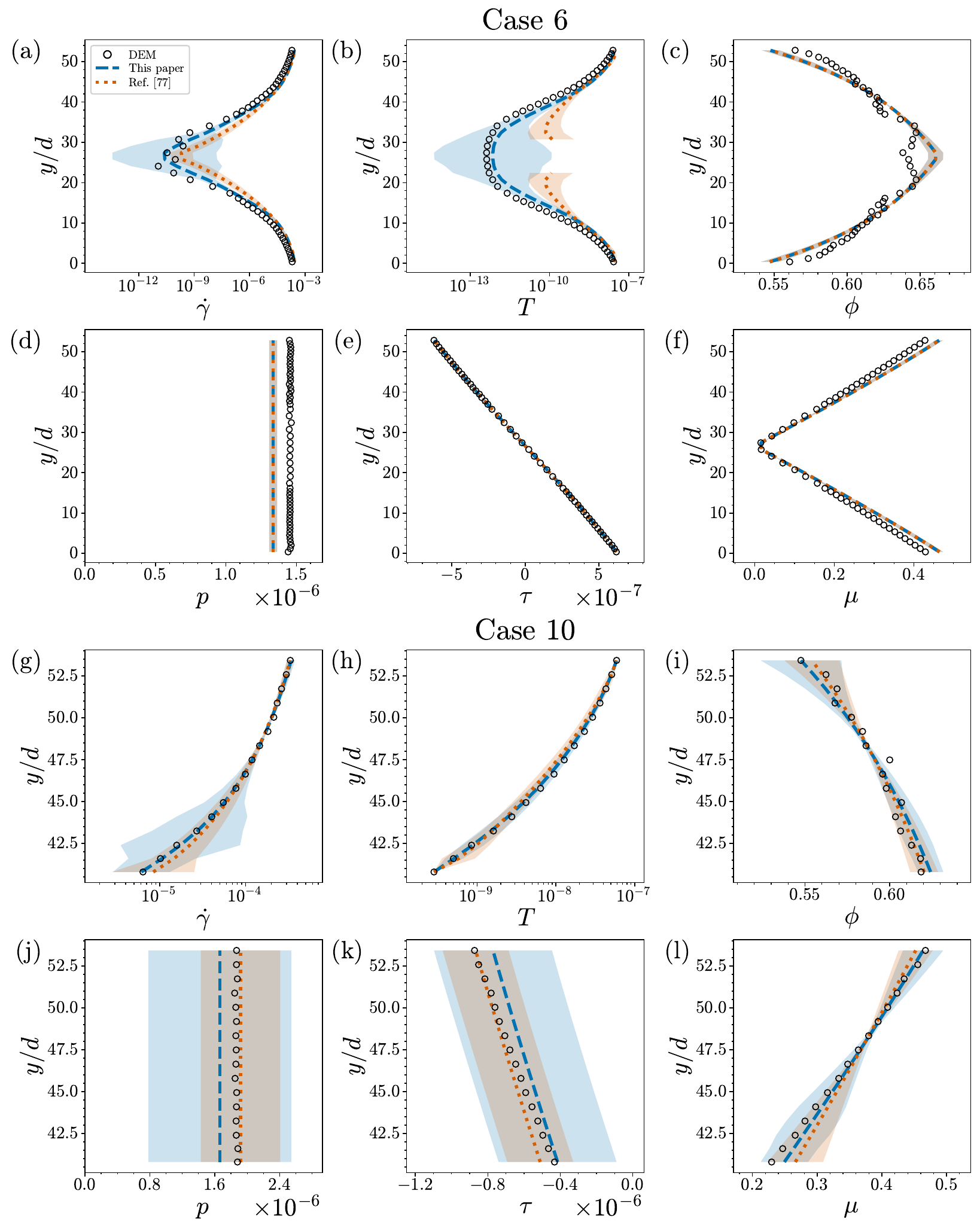}
\caption{\label{fig:model_solutions_sensitivity}The solutions for the models of this paper and Ref.~\cite{kamrin2014nonlocal} with error estimates. }
\end{figure}

We solve the system of equations using our measured $\mu(\phi)$ and $\mu(I,\Theta)$, and for three models of the temperature equation: this paper's, NGF~\cite{kamrin2014nonlocal}, and HE~\cite{irmer2025granular}. For this paper we use $\tilde\kappa(\Theta)$ and $\tilde\chi(I,\Theta)$ defined above. For NGF, we replace Eq.~\ref{eqn:generic_temperature_equation} with Eq.~\ref{eqn:NGF_full}, and infer $T$ from the $\mu(I,\Theta)$ relation. For HE, we replace Eq.~\ref{eqn:generic_temperature_equation} with Eq.~\ref{eqn:heat_equation}. The model of this paper and NGF use the parameters in Tbl.~\ref{tbl:scaling_fit_parameters}, which are calibrated to our DEM data. For HE, our DEM-calibrated parameters resulted in a poorly conditioned equation yielding solutions deviating extremely far from DEM, which is consistent with the observation of huge uncertainty in their values. Instead, we utilize the parameters $D_\text{HE}=1$, $A_\text{HE}=0.15$, $B_\text{HE}=1$, and $a_\text{HE}=1.5$ selected by Ref.~\cite{yuan2026toward}, which would correspond to a small subset of our data. 

We assemble a system of six first-order ODEs, where we introduce a term $M$ that integrates to give the packing fraction. Notating $df/dy=f'$, we have for this paper
\begin{subequations}
\begin{equation}
    u_x'=\dot\gamma,
\end{equation}
\begin{equation}
    \tau' = -\rho\phi G_x, 
\end{equation}
\begin{equation}
    p' = \rho\phi G_y,
\end{equation}
\begin{equation}
    T'  = -\frac{Q_y}{\kappa},
    \label{eqn:Tprime}
\end{equation}
\begin{equation}
    Q_y' = \tau\dot\gamma - \chi,
    \label{eqn:Qprime}
\end{equation}
\begin{equation}
    M' = \phi,
\end{equation}
\end{subequations}
with boundary conditions $u_x(y_b)=U_b$, $u_x(y_t)=U_t$, $T(y_b)=T_b$, $T(y_t)=T_t$, $M(y_b)=0$, $M(y_t)=\int_{y_b}^{y_t}\phi dy$. We solve the system with \verb|scipy.integrate.solve_bvp|, utilizing the DEM profiles as initial guesses and inverting $\phi$ from $\mu(\phi)$. Note that (apart from the initial guesses) we are not supplying the stress fields nor their boundaries from DEM, a common practice that, as pointed out in Ref.~\cite{debnath2022comparison} might minimize model inaccuracies. We also impose the viscosity regularizer of Ref.~\cite{kim2020power}. For NGF, we replace Eqns.~\ref{eqn:Tprime} and~\ref{eqn:Qprime} with, for fluidity $g=\dot\gamma/\mu$,
\begin{subequations}
\begin{equation}
    g' = Q_\text{NGF},
\end{equation}
\begin{equation}
    Q_\text{NGF}'= \frac{1}{A_\text{NGF}^2d^2}\left((\mu_2-\mu_s)\frac{\mu_s-\mu}{\mu_2-\mu}g + \frac{\mu_2-\mu_s}{I^\star}\sqrt{\frac{\rho d^2}{p}}\mu g^2 \right),
\end{equation}
\end{subequations}
and infer the boundary conditions on $g$ from the $T$ boundary conditions, and the $T$ solution field from $g$ (with clipping for predicted values outside of the fitted function ranges). For HE, we replace Eqns.~\ref{eqn:Tprime} and~\ref{eqn:Qprime} with
\begin{subequations}
\begin{equation}
    \Theta' = Q_\text{HE}
\end{equation}
\begin{equation}
    Q_\text{HE}' = \frac1{D_\text{HE}d^2}(B_\text{HE}\Theta-A_\text{HE}I^{a_\text{HE}}),
\end{equation}
\end{subequations}
and infer the boundary conditions on $\Theta$ from the $T$ boundary conditions, and the $T$ solution field from $\Theta$. 

Figure~\ref{fig:model_solutions} plots example solutions, comparing the three models. All three models are reasonably accurate for the first case, but HE greatly loses accuracy for another case. Error estimates for this paper and NGF are shown in Fig.~\ref{fig:model_solutions_sensitivity}, excluding HE because its parameters were arbitrarily chosen. To generously estimate error, we removed a random 70\% of the data before calculating parameters, determined the solutions, and repeated 100 times to obtain a standard deviation for each $y$-location of every solution field, demonstrating that our model is robust.

\end{document}